\documentclass[aps,prb,preprint,amssymb]{revtex4}
\usepackage{epsfig}
\usepackage{dcolumn}% Align table columns on decimal point
\usepackage{amsmath}
\usepackage{rotating}
\begin{document}
\renewcommand{\thefootnote}{\fnsymbol{footnote}} 
\title{Quantum contributions in the ice phases: the path to a new empirical model for water -- TIP4PQ/2005}
\author{Carl McBride$^a$} 
\author{Carlos Vega$^a$} 
\email[]{E-mail: cvega@quim.ucm.es}
\author{Eva G. Noya$^a$}
\affiliation{$^a$Dept. Qu\'{\i}mica-F\'{\i}sica I, Fac. de Ciencias Qu\'{\i}micas, Universidad Complutense de Madrid, 28040 Madrid, Spain}
\author{Rafael Ram\'irez$^{b}$}
\affiliation{$^b$Instituto de Ciencia de Materiales, CSIC, Campus de Cantoblanco, 28049 Madrid, Spain}
\author{Luis M. Ses\'e$^c$} 
\affiliation{$^c$Dept. Ciencias y T\'ecnicas Fisicoqu\'{\i}micas, Fac. de Ciencias, UNED, Paseo Senda del Rey 9, 28040 Madrid, Spain}
\date{\today}
\begin{abstract}
\noindent  
With a view to a better understanding of the influence of atomic quantum delocalisation effects
on the phase behaviour of water, path integral simulations have been undertaken for almost all 
of the known ice phases using the TIP4P/2005 model, in conjunction with  the rigid rotor propagator proposed by  M{\"u}ser and Berne 
[Phys. Rev. Lett. {\bf 77}, 2638 (1996)].
The quantum contributions then being known, a new empirical model of water
is developed (TIP4PQ/2005) which reproduces, to a good degree, a number of the 
physical properties of the ice phases, for example densities, structure and relative stabilities.
\end{abstract}
\maketitle
\section{Introduction}
\label{intro}
``Water, water, every where..." goes the poet Samuel Taylor Coleridge's {\it The Rime of the Ancient Mariner},
which provides a magnificent r\'esum\'e of our reason for studying this ubiquitous material.
Many volumes have been written about water and ice (to cite just a few
\cite{bookStruPropWater,bookPhysIce,book_FFranks_Water_Matrix_Life,book_Ball_Life_Matrix_Water,book_WaterInterfBulkFaraday}),
and a good deal more await writing, before we fully understand  this enigmatic molecule.

Currently the point has been reached where many properties, 
including the global phase diagram of water and the ice phases, 
can be reproduced qualitatively (and in some cases, quantitatively) 
using little more than a simple empirical model \cite{PRL_2004_92_255701}.
However, there are several aspects of water where our knowledge, and thus our understanding,
is far from complete.
One such aspect is the high pressure/temperature  region of the phase diagram, where the precise location of the melting curves
is still yet to be agreed upon due to the difficult nature of the experiments.
For example, it is an open question as to whether water becomes super-ionic in this region \cite{PRL_2005_94_125508,PNAS_2008_105_14779}.
In one of the ice polymorphs, ice X, the notion of a water molecule even becomes lost, the protons 
being shared equally between oxygen atoms \cite{N_1998_392_00258,N_1999_397_00503}.
The low temperature region of the phase diagram is also extremely interesting, where a 
host of `anomalous' or atypical trends are also present.
Examples are  
the well known maximum in density at 3.984 Celsius, a minimum in the isothermal compressibility at 46.5 Celsius, 
and an unusual variation of the diffusion coefficient with pressure. 
These trends are especially apparent in super-cooled water where 
one can also find a minimum in both the density \cite{PNAS_2007_104_09570} 
and a dynamic transition to Arrhenius behaviour for the diffusion coefficient \cite{PNAS_2005_102_16558,PNAS_2007_104_09575}.
It has been suggested that 
many of the anomalous properties of water at low temperatures could be understood
by an hypothesised second
critical point \cite{N_1992_360_00324,PT_2003_56_06_0040,JPCM_2003_15_R1669} 
buried deep within ``no-mans land" \cite{PCCP_2000_02_1551}, a region of the phase diagram inaccessible 
to experiment. 
If this is so, it would go a long way to explaining another feature of water;  its capacity to form several
amorphous phases (glasses) at low temperatures.

In elucidating the origin of these anomalies, computer simulations have played a prominent role,
for example their part in the proposal of a  second critical point in water \cite{N_1992_360_00324,PA_1994_205_0122} using a 
simple empirical model.
Classical computer simulations do, however, have their limitations.
There are certain systems, water being one of them, where quantum effects are 
significant \cite{PRL_2008_101_017801,JPCB_2009_113_05702}.
As an example, let us examine the difference in temperature between the 
melting point and the temperature of maximum density.
For H$_2$O this amounts to 3.984K, whereas for D$_2$O it is 7.365K.
From the point of view of the Born-Oppenheimer approximation
the potential energy surface (PES) is independent of the isotope considered.
Thus the different behaviour of these isotopes is due to how the molecules
react to this PES. This is known as an  atomic quantum delocalisation effect.
In this particular case the origin of the differences, both structural and dynamical,  is in good part due to the 
quantum nature of the hydrogen protons and the strength of the hydrogen bond.
Another example is the self-diffusion coefficient, which  increases by more than 50\% 
in a quantum system with respect to classical molecular dynamics simulations \cite{JCP_2006_125_054512,JCP_2005_123_154504}.

The overall structure of water is that of an asymmetric top, which is to say that all 
three principal moments of inertia are distinct. 
What is particularly interesting is that since hydrogen is the 
lightest atom,  the rotational moments of inertia  
are small enough to show marked 
quantum behaviour. Thus water has significant quantum effects even at room temperature.
The importance of these quantum effects increases as the temperature is lowered.
For the ice phases these effects are expected to be significant, especially at 
77K where many experiments on ice are frequently performed using liquid nitrogen.
Thus far there has been relatively little work on these effects for ice, 
and almost all of the work that has been published has focused on ice I$_h$ 
\cite{JCP_1996_104_00680,JCP_2005_123_144506,JCP_2006_125_054512,JPCC_2008_112_00324}.
The objective of this publication is to quantify the size of these
effects in all of the ice phases, apart from that of ice X, which cannot be described 
by the rigid models used in this work.

These  atomic quantum delocalisation effects will be studied using the empirical TIP4P/2005 model \cite{JCP_2005_123_234505}.
Over the last few years a number of the present authors have undertaken extensive simulation 
studies examining the performance of a number of commonly used models for water,
in particular the TIP3P, TIP4P, TIP5P and SPC/E models \cite{FD_2009_141_0251}. The principal findings 
have been that the TIP3P \cite{JCP_1983_79_00926}, TIP5P \cite{JCP_2000_112_08910}  and SPC/E \cite{JPC_1987_91_06269} 
models experience difficulties when it comes
to describing the global phase diagram of water and the ice phases. However, 
the TIP4P model does indeed provide a qualitatively correct phase diagram.
Based on this finding, the TIP4P model was re-parameterised in order to 
improve the quantitative representation, leading to the TIP4P/2005 model \cite{PCCP_2009_11_0556}.
It has since been found that this model also provides a good description of  the 
maximum in density of liquid water and its variation with pressure \cite{MP_2009_107_0365}, 
of the compressibility minima \cite{MP_2009_107_0365}, 
the surface tension \cite{JCP_2007_126_154707}, 
the vapour liquid equilibria \cite{JCP_2006_125_034503},
the critical properties \cite{JCP_2006_125_034503}, 
the equation of state at high pressures \cite{FD_2009_141_0251},
the diffusion coefficient \cite{FD_2009_141_0251},
and the viscosity \cite{FD_2009_141_0251}. 

That said, the model was parameterised for classical simulations, so the 
introduction of  atomic quantum delocalisation effects, although improving the qualitative 
description, will cause a deterioration in  the quantitative description.
In the first stage of 
this research we shall analyse the impact of  atomic quantum delocalisation effects on the properties of 
the ice phases using this potential. That will elucidate  where, and how,  atomic quantum delocalisation
effects modify the properties of water with respect to the classical limit.  
These differences then known, we provide a re-parameterised version of the TIP4P/2005
model which we shall call the TIP4PQ/2005 model, the Q indicating that this 
model is suitable for quantum simulations.
As was pointed out by Morse and Rice \cite{JCP_1982_76_00650} as well as by Whalley ``...effective potentials that are used to simulate water
ought to be tested on the many phases of ice before being treated as serious representations of liquid water" \cite{JCP_1984_81_04087}. 

\section{Methodology}
\label{exper}
Simulations were performed using the path integral
formulation, which permits us to study the quantum effects 
related to the finite mass of the atoms
(in many quantum chemistry calculations, the electrons are treated 
as being quantum, however the nuclei are treated as classical point masses).
A particularly elegant technique for studying quantum effects in many body systems is that of path-integral Monte Carlo (PIMC).
There are many good introductions concerning PIMC in the literature 
\cite{ebook_FeynmanHibbs,JCP_1981_74_04078,NATO_ASI_C_293_0155_photocopy,RMP_1995_67_00279,AllenTildesleyChp10},
here we shall focus on the aspects most pertinent to the simulations we have performed.

Water is, of course, a flexible molecule. 
For path integral simulations one generally requires the number of Trotter slices, $P$, to be  \cite{JCP_2008_129_024105}
\begin{equation}
P > \frac{ \hbar \omega_{\mathrm{max}}} {k_B T}
\end{equation}
where $ \omega_{\mathrm{max}} $ is the `fastest' frequency present in the 
system in question. 
In water the intramolecular vibrations are of the order of
$ \omega_{\mathrm{max}} / 2\pi c \approx 4000 ~\mathrm{cm}^{-1} $
which leads to $P>20$. Using the rigid body approximation for water 
the fastest motion now becomes the libration, with a frequency of 
$ <900 ~\mathrm{cm}^{-1} $, thus reducing $P$ to around 5-6.
This represents a substantial reduction in the computational overhead 
associated with traditional PIMC calculations (although new techniques have 
recently been developed by Manolopoulos et al. to increase the efficiency of flexible molecule PIMC \cite{JCP_2008_129_024105}). 
It must be said that by choosing to use a rigid model, one precludes the
ability to study some aspects of water, such as the high frequency region of the infra-red adsorption spectrum \cite{PNAS_2005_102_06709,JCP_2008_129_074501}.
The infra-red spectrum of water and  ice can be divided up into two distinct regions.
Above  $\approx 900  ~\mathrm{cm}^{-1} $ one has the contribution associated with  the intramolecular degrees of freedom of bending and stretching.
Below $\approx 900  ~\mathrm{cm}^{-1} $, as previously mentioned,  one has the section that corresponds to translational and librational movements, and are mostly due to inter-molecular forces.
Quantum contributions to the Helmholtz energy $(A)$ within a perturbative treatment for a rigid asymmetric top are given by \cite{MP_1979_38_1875_nolotengo}:
\begin{equation}
\frac{A-A_{C1}}{N_m} =  \frac{\hbar^2}{24(k_BT)^2} \left[ \frac{\left< F^2 \right>}{M} +  \frac{\left< \Gamma_A^2 \right>}{I_A} +   \frac{\left< \Gamma_B^2 \right>}{I_B}  +  \frac{\left< \Gamma_C^2 \right>}{I_C}   \right] - \frac{\hbar^2}{24} \sum_{\mathrm {cyclic}} \left( \frac{2}{I_A} - \frac{I_A}{I_BI_C}\right) + {\mathcal {O}} (\hbar^4)
\end{equation}
A good proportion of the quantum effects in water are due to the
strength of the hydrogen bond, along with a particularly small inertia tensor.
It is this that lends importance to the torque $(\Gamma)$ terms found in
the above equation, which results in the appearance of the librational band.
In contrast, this region for a molecule such as SO$_2$, where no such hydrogen bonding is present,
is far less important.
By using the path integral formulation for a rigid model we shall be studying atomic quantum delocalisation effects in the influential region encountered below  
 $\approx 900  ~\mathrm{cm}^{-1} $. 
In a study of the phonon density of states for ice I$_h$ Dong and Li \cite{PB_2000_276_0469} 
showed that the rigid TIP4P model does a reasonable job of reproducing this 
low frequency section of the spectrum.
Even given the fact that intramolecular effects are important, it is surely the case that a
rigid body path-integral study is more physically realistic than a purely classical
study, which neglects all atomic quantum delocalisation effects.
Such an approach has been adopted in a number of studies, using for example the SPC/E model \cite{JCP_2005_123_154504}.
In view of this, and given the success that the TIP4P/2005 model has had in describing 
the ice phases classically, the rigid TIP4P/2005 model is the natural candidate for a preliminary study 
of  atomic quantum delocalisation effects in ices.
Given that the TIP4P/2005 model is a rigid asymmetric top, we shall first present 
the path integral description of a rigid rotor.
\subsection{Path integrals for a rigid molecule}
\label{one_rotor}
The coordinates used to describe a rigid molecule are
$\mathbf{r}_1\mathbf{\Omega}_1$, where $\mathbf{r}_1$ represents  the centre of mass and 
$\mathbf{\Omega}_1=(\phi_1, \theta_1, \chi_1)$ represents the Euler angles 
that fix the molecule orientation.
The Hamiltonian of a rigid asymmetric rotor
can be written in the form \cite{JPCM_1999_11_0R117}:
\begin{equation}
\hat{H}_1 = \hat{T}^{\mathrm {tra}} + \hat{T}^{\mathrm {rot}}+ \hat{U} \, ,
\end{equation}
where $\hat{T}^{\mathrm {tra}}$ represents the kinetic energy operator
associated to the centre of mass translation, $\hat{U}$ appears as a
potential energy operator that is a function of the
coordinates $\mathbf{r}_1 \mathbf{\Omega}_1$, and
the rotational kinetic energy operator is given by \cite{JPCM_1999_11_0R117}:
\begin{equation}
\hat {T}^{\mathrm {rot}} = \sum_{i=1}^{3} \frac{\hat{L}_i^2}{2I_i} \, ,
\end{equation}
where $\hat{L}_i$ are the components of the angular momentum operator and  $I_i$ are the moments of inertia of the molecule referred to
its fixed body frame. We assume, without loss of generality,
that the moment of inertia tensor is diagonal in the chosen
fixed body frame.

In the path integral formulation, the partition function, $Q_1$, 
of a rigid molecule may be expressed by a factorisation of the density matrix 
into $P$ factors, so that each quantum particle is described by a ring 
of $P$ replicas or `beads',
\begin{eqnarray}
Q_1(\beta) &=& \lim_{P \rightarrow \infty} 
\int \ldots \int \prod_{t=1}^P \mathrm{d}\mathbf{r}_1^t  \mathrm{d}\mathbf{\Omega}_1^{t}
                 \prod_{t=1}^P \rho_1^{t,t+1} (\beta/P) \, ,
\end{eqnarray}
where $\beta=1/k_BT$ is the inverse temperature, and the propagator
$\rho_1^{t,t+1}$ is approximated by \cite{JPCM_1999_11_0R117}:
\begin{equation}
\rho_1^{t,t+1}(\beta/P) \approx \left< \mathbf{r}_1^t \mathbf{\Omega}_1^t \vert \exp \left[ -\beta \hat{U} /2P \right] \exp \left[ -\beta (\hat{T}^{\mathrm{tra}} + \hat{T}^{\mathrm{rot}}) /P \right] \exp \left[ -\beta \hat{U} /2P \right]  \vert  \mathbf{r}_1^{t+1} \mathbf{\Omega}_1^{t+1} \right> . 
\end{equation}
The propagator satisfies the cyclic condition that bead $P+1$ corresponds
to bead $1$.
This rigid molecule propagator is built up of three factors, 
a potential energy component,
a translational component, and a rotational component:
\begin{equation}
\rho_1^{t,t+1}(\beta/P) \approx \rho^{t,t+1}_{\mathrm{pot},1} \rho^{t,t+1}_{\mathrm{tra},1} \rho^{t,t+1}_{\mathrm{rot},1}. 
\end{equation}
The potential energy component is given by  \cite{JPCM_1999_11_0R117}
\begin{equation}
\rho^{t,t+1}_{\mathrm{pot},1} = \exp \left[ - \frac{\beta}{2P} 
             \left( U^t +  U^{t+1} \right) \right]
\, ,
\end{equation}
where $U^t= U ( \mathbf{r}_1^t {\bf \Omega}_1^t)$ is the potential 
energy of the replica $t$ of the molecule.
The translational component is given by  \cite{JPCM_1999_11_0R117} 
\begin{equation}
\rho^{t,t+1}_{\mathrm{tra},1} = \left< \mathbf{r}_1^t \vert \exp \left( -\beta \hat{T}^{\mathrm{tra}} /P \right) \vert \mathbf{r}_1^{t+1} \right>
= \left( \frac{MP}{2 \pi \hbar^2 \beta} \right)^{3/2}  \exp \left[ -  \frac{MP}{2 \hbar^2 \beta} ( \mathbf{r}_1^{t} -  \mathbf{r}_1^{t+1})^2 \right],
\end{equation}
where $M$ is the total mass of the rigid molecule.
The two previous equations are well known and are commonly used as the
so-called primitive approximation in 
path integral studies of simple fluids.
The rotational propagator between  $t$ and $t+1$  is given by  \cite{JPCM_1999_11_0R117}:
\begin{equation}
\rho^{t,t+1}_{\mathrm{rot},1} =  \left<  {\bf\Omega}_1^t \vert \exp \left[ -\beta \hat{T}^{\mathrm{rot}} /P \right] \vert {\bf\Omega}_1^{t+1} \right>.
\end{equation}
In an important piece of work  M{\"u}ser and Berne \cite{JPCM_1999_11_0R117,PRL_1996_77_002638} have shown that the rotational 
contribution to the propagator 
between the replicas $t$ and $t+1$ of a rigid molecule $i$ is exactly given by
\begin{equation}
\rho_{{\mathrm{rot}},i}^{t,t+1}( \tilde{\theta}_{i}^{t,t+1}, \tilde{\phi}_{i}^{t,t+1} + \tilde{\chi}_{i}^{t,t+1})   = 
\sum_{J=0}^{\infty} \sum_{M=-J}^J \sum_{\tilde{K}=-J}^J  
f_{i,J,M,\tilde{K}}^{t,t+1}   \exp \left( - \frac{\beta E^{JM}_{\tilde{K}} }{P}   \right) 
\end{equation}
where 
\begin{equation}
f_{i,J,M,\tilde{K}}^{t,t+1} = 
\frac{2J+1}{8\pi^2}  d^J_{MM} (\tilde{\theta}_{i}^{t,t+1}) 
\cos[M(\tilde{\phi}_i^{t,t+1}+\tilde{\chi}_i^{t,t+1})] 
|A_{\tilde{K}M}^{(JM)}|^2 
\end{equation}
Here $d^J_{MM} (\tilde{\theta}_i^{t,t+1})$ are Wigner
functions and $|A_{\tilde{K}M}^{(JM)}|$ are the coefficients of
the expansion of the eigenstates of the asymmetric top in 
a basis formed by the eigenstates of the symmetric top. $E_{\tilde{K}}^{(JM)}$
are the eigenvalues of the energy of the asymmetric top. The quantum numbers
$J$ and $M$ provide the values of the total angular momenta of the asymmetric top
and the value of its $z$ component in the laboratory frame. 
The number $\tilde{K}$ is 
not a true quantum number, in the sense that it does not provide the value of any physical observable,
but rather is an index used to label the $(2J+1)$ energy levels that are obtained
for each value of $J$. 
The angles $\tilde{\theta}^{t,t+1}_i$, $\tilde{\phi}^{t,t+1}_i$ and 
$\tilde{\chi}^{t,t+1}_i$ are the Euler angles of the replica $t+1$ 
of molecule $i$ expressed in the body frame fixed in the 
replica $t$ of the same molecule $i$. 
Note that the rotational propagator depends solely on two variables,
$\tilde{\theta}^{t,t+1}_i$ and  $\tilde{\phi}^{t,t+1}_i +\tilde{\chi}_i^{t,t+1}$. 
Obviously to determine the value of the rotational propagator one must first 
determine the $(2 J +1 )$ energy levels of the asymmetric top for each value of $J$. 
This can be obtained from the $(2 J +1)$ eigen-values, $E_{\tilde{K}}^{(JM)}$, of the matrix given in Ref. \onlinecite{book_Zare_AngMom}.
The coefficients $|A_{\tilde{K}M}^{(JM)}|$ are the eigen-vectors associated with these eigen-values.
It is computationally convenient to calculate the rotational propagator
$ \rho_{{\mathrm{rot}},i}^{t,t+1}( \tilde{\theta}_{i}^{t,t+1}, \tilde{\phi}_{i}^{t,t+1} + \tilde{\chi}_{i}^{t,t+1})$ 
for a grid of values of the angles $\tilde{\theta}^{t,t+1}_i$ and  $\tilde{\phi}^{t,t+1}_i +\tilde{\chi}_i^{t,t+1}$ for each value of $\beta/P$ to be used,  
and save this data prior to the simulations. 
The value of the rotational propagator for any given $\tilde{\theta}^{t,t+1}_i$ and  $\tilde{\phi}^{t,t+1}_i +\tilde{\chi}_i^{t,t+1}$ 
can then be estimated using a linear interpolation algorithm from this tabulated data.

\subsection{Path integrals for an ensemble of rigid molecules}
\label{ensemble_rotor}
Once the translational and rotational propagators are known
for a rigid molecule one can 
calculate the partition function for a set of interacting molecules. 
Let us assume that we shall be using a pair-wise potential  $u_{ij}$
such that the potential energy of the replica $t$ of the system is 
\begin{equation}
 U^t
= \sum_i \sum_{j>i} u_{ij} (\mathbf{r}_i^t,\mathbf{r}_j^t, {\bf \Omega}_i^t,{\bf \Omega}_j^t).
\label{Ut}
\end{equation}
Now the canonical 
partition function, $Q_N$, of an ensemble of $N$ molecules described with $P$ beads
is given by:
\begin{eqnarray}
Q_{N}(\beta) & \approx & \frac{1}{N!}\left( \frac{MP}{2\pi\beta \hbar^2 } \right)^{3NP/2} \int \ldots \int \prod_{i=1}^N \prod_{t=1}^P d\mathbf{r}_i^t d{\bf\Omega}_i^t
\times \nonumber \\
& &
\exp \left( - \frac{MP}{2\beta \hbar^2} 
\sum_{i=1}^N \sum_{t=1}^P\left( \mathbf{r}_i^t - \mathbf{r}_i^{t+1}\right)^2 
- \frac{\beta}{P} \sum_{t=1}^P U^t \right)
\prod_{i=1}^N \prod_{t=1}^P \rho_{{\mathrm{rot}},i}^{t,t+1} \, .
\label{partition_function}
\end{eqnarray}
As can be seen in Eqs. (\ref{Ut}) and (\ref{partition_function}), 
each replica $t$ of molecule
$i$ interacts: {\it (a)} with the molecules that have the same index 
$t$ via the intermolecular potential $u_{ij}$; {\it (b)}
with replicas $t-1$ and $t+1$ of the same molecule $i$ via a harmonic
potential whose coupling parameter depends on the mass of the molecules, $M$,
and on the inverse temperature $\beta$; 
and {\it (c)} with replicas $t-1$ and $t+1$ of the same molecule
through the terms $\rho_{{\mathrm{rot}},i}^{t-1,t}$ and $\rho_{{\mathrm{rot}},i}^{t,t+1}$  which incorporate the quantisation
of the rotation, which in turn  depends on the relative orientation
of replica $t$ with respect to $t-1$, and  $t+1$ with respect to  $t$. 

Let us define an energy $U'$ as: 
\begin{equation}
U' = 
   \frac{MP}{2\beta^{2} \hbar^2} 
\sum_{i=1}^N \sum_{t=1}^P\left( \mathbf{r}_i^t - \mathbf{r}_i^{t+1}\right)^2 
+ \frac{1}{P} \sum_{t=1}^P U^t \, ,
\label{Uprime}
\end{equation}
and the total orientational propagator $P_{\mathrm {rot}}$ as: 
\begin{equation}
P_{\mathrm {rot}}= 
\prod_{i=1}^N \prod_{t=1}^P \rho_{{\mathrm{rot}},i}^{t,t+1}  \, .
\end{equation}
Within a  Monte Carlo simulation one generates a new configuration starting from 
a previous configuration. The probability of accepting this new configuration, $p_{accept}$, 
is given by 
\begin{equation}
  p_{accept}=  \min \left[ 1, \exp \left(- \beta  ( U'_{new} - U'_{old} ) \right) \frac{P^{new}_{{\mathrm{rot}}}}{P^{old}_{{\mathrm{rot}}}} \right] \, .
\end{equation}
It is worthwhile making  two observations about the orientational propagator between a pair of contiguous beads 
$ \rho_{{\mathrm{rot}},i}^{t,t+1}$.
Firstly, it must be positive in order to 
be used in the Metropolis acceptance criteria, 
which is indeed the
case. 
Secondly, the maximum in the orientational
propagator is achieved when $\tilde{\theta}=0$ and $\tilde{\phi} + \tilde{\chi} = 0$.
It is found that at high enough temperature
the propagator decays to zero relatively quickly as the values of 
$\tilde{\theta}$ and $\tilde{\phi} + \tilde{\chi}$ 
increase. 
The orientational propagator can also be expressed as an auxiliary 
energy by defining
$u_{i,aux}$ such that 
\begin{equation}
  u_{i,aux}^{t,t+1} = -\frac{1}{\beta} \ln  \rho_{{\mathrm{rot}},i}^{t,t+1}  
\end{equation} 
$u_{i,aux}$ has a minimum at  $\tilde{\theta}=0$ and $\tilde{\phi} + \tilde{\chi} = 0$ and 
increases quickly as a function of the
variables $\tilde{\theta}$ and $\tilde{\phi} + \tilde{\chi}$.
$P_{\mathrm {rot}}$ can now be written as
\begin{equation}
 P_{\mathrm {rot}}= \exp(-\beta U_{\mathrm {aux}} ) = \exp \left( - \beta \sum_{i=1}^N \sum_{t=1}^P  u_{i,aux}^{t,t+1} \right).
\end{equation}
Using this auxiliary energy the Metropolis criteria can be now written as :
\begin{equation}
   p_{accept}= \min \left[1, \exp\left(- \beta \left( (U'_{new} + U_{\mathrm {aux},new} ) - ( U'_{old} + U_{\mathrm{aux},old})\right)\right)\right] \, .
\end{equation} 
This expression helps us to clarify  the role of the orientational propagator; it can be viewed as a potential
that forces two contiguous beads, $t$ and $t+1$, to adopt similar orientations (this corresponds to the minimum of the
auxiliary potential) with an energetic penalty when they adopt different orientations. 
This is analogous to the role played by the harmonic springs connecting the centre of masses of the molecules in Eq. (\ref{Uprime}). 

The internal energy can now be calculated from:
\begin{equation}
E = - \frac{1}{Q_{N}} \frac{\partial Q_{N}}{\partial \beta } \, .
\end{equation}
It can be shown that substituting the value of the canonical partition
function in this expression results in 
\begin{equation}
E =  K_{\mathrm {tra}} + K_{\mathrm {rot}} + U  \, ,
\end{equation}
where :
\begin{eqnarray}
K_{tra} & = &\frac{3NP}{2\beta } - \left\langle \frac{MP}{2\beta^2 \hbar^2}
	\sum_{i=1}^N \sum_{t=1}^P ({\bf r}_i^t - {\bf r}_i^{t+1})^2 
        \right\rangle ,  \nonumber \\
K_{rot}  & =  & \left\langle \frac{1}{P} \sum_{i=1}^N \sum_{t=1}^P 
\frac{ \sum_{J=0}^{\infty } \sum_{M=-J}^{J} \sum_{\tilde{K}=-J}^J f_{J,M,\tilde{K}}^{i,t,t+1}
{\tilde E}_{\tilde K}^{JM} \exp{\left[ -\frac{\beta}{P} \tilde{E}_{\tilde{K}} ^{JM} \right]}}
{ \rho_{rot,i}^{t,t+1}  }
\right\rangle ,  \nonumber\\ 
U  & =   & \left\langle  \frac{1}{P}  \sum_{t=1}^P U^t 
   \right\rangle  \, .
\label{energy}
\end{eqnarray}
As with the rotational propagator, the numerator of $K_{rot}$ in Eq. \ref{energy} 
was calculated prior to the simulations for a grid of the variables 
$\tilde{\theta}$ and  $\tilde{\phi} +\tilde{\chi}$
and subsequently saved in tabular form. 

When performing simulations of solids it is more convenient to perform the 
simulations in the $NpT$ ensemble.
The partition function for the $NpT$ ensemble can be calculated using: 
\begin{equation}
Q_{NpT}= A \int dV \exp(-\beta pV) Q_{N}
\end{equation}
where $A$ is a constant with units of inverse volume that makes $Q_{NpT}$
dimensionless. Its value affects the Helmholtz energy function, but not the configurational
properties. 
\subsection{Simulation details}
In this work path integral Monte Carlo simulations are undertaken for the TIP4P/2005 model 
for fourteen of the fifteen known ice phases.
One of the most important variables when it comes to path integral simulations is the 
number of Trotter slices, or beads, ($P$) employed. If $P=1$ then the simulation is classical.
As $P \rightarrow \infty$ then the quantum simulation becomes exact.
Given the isomorphism between Trotter slices and the number of component `beads' in a ring polymer \cite{JCP_1981_74_04078}, one can easily see
that the time required for a simulation scales with the number of Trotter slices used.
For flexible models of water at 300K a typical number of slices is about $P=24$ \cite{JCP_1997_106_02400,JCP_2001_115_07622,PRE_2005_71_041204}.
However, if a rigid model is employed, the number of
Trotter slices required can be reduced by about
a factor of five \cite{JCP_2001_115_10758,JCP_2005_123_144506}.
Previous studies for a rigid model of  water at 300K found that a value of $P=5$ provides good convergence \cite{JCP_1996_104_00680,JCP_2005_123_144506}.
Thus in this work  the number of Trotter slices times the temperature was maintained at $PT\approx 1500$.  
For the lowest considered temperature (77K) this corresponds to 20 beads.
When computing the asymmetric top eigen-energies and eigenvectors of water
the OH distance and the H-O-H bond angle
of the TIP4P/2005 model were used, which corresponds to the gas phase geometry of real water.
The principal moments of inertia are computed using this geometry along with  the masses of the
hydrogen and oxygen atoms. Although the model has the negative charge on the site M,
this site is massless and therefore it is only used to compute the potential energy of the
system.

In this work two models of water are studied, the TIP4P/2005 model \cite{JCP_2005_123_234505}
and a re-parameterisation, which we shall call the TIP4PQ/2005 model,  to `compensate' for quantum effects.
The parameters for both of these models are given in Table \ref{tip4p_2005_parameters}.
The only difference between these models 
is an increase in the charges on the hydrogen sites by 0.02e, along with a corresponding increase in the
charge on the oxygen site.
For both models the Lennard-Jones potential was truncated at 8.5{\AA} and long-range corrections were included.
The TIP4P/2005 model has been designed to be used with Ewald summations \cite{PRSLA_1980_373_0027,PRSLA_1980_373_0057}
which is a well known technique to treat the long range electrostatic interactions.
Ewald summation is more appropriate than the reaction field method when it
comes to the  simulation of solid phases.
The real part of the Coulombic potential was truncated at 8.5\AA.

The configurational space of the quantum system was
sampled using a Monte Carlo code with four distinct types of
trial moves: the displacement of a single bead of one molecule,
rotation of a single bead of one molecule, translation of a whole
ring, and rotation of all of the replicas of one molecule. A Monte Carlo cycle
is defined as $N$  Monte Carlo moves, where the probability of
attempting a translation or a rotation of a single bead is 30\% each
and the probability of attempting a translation of a whole ring
or rotating all the replicas of a ring is 20\% each.
The maximum displacement or rotation in each type of movement was adjusted to
obtain a 40\% acceptance probability. When simulations were
performed in the $NpT$ ensemble,  besides the $N$ particle
trial moves, one Monte Carlo cycle also includes an attempt to change the  volume of the simulation box.
The maximum volume change was adjusted so as to obtain a 30\%
acceptance probability.
In general the simulations consisted of 30,000 Monte Carlo
equilibration cycles, followed by a further 100,000 cycles
for the accumulation of run averages. 
The number of molecules used in each of the
phases are given in Table \ref{table_ice_phases_1}.
For the proton disordered ice phases the positions of the hydrogen atoms were 
generated in such a way as to produce a system that satisfies the so-called Bernal-Fowler ice rules \cite{JCP_1933_01_00515,JACS_1935_57_02680},
and whose dipole moment as close as possible to  zero.
This was achieved using the algorithm proposed by Buch {\it et al.} \cite{JPCB_1998_102_08641,JCP_2004_121_10145}.

As mentioned, all simulations were performed in the isothermal-isobaric ($NpT$) ensemble. 
The implementation of the $NpT$ ensemble in PIMC has
already been discussed in previous works \cite{JCP_1989_90_05644,PRB_1995_51_002723}.
It is important to note that the Monte Carlo volume moves should be performed
anisotropically, in order to allow the simulation box to `relax' and obtain
the true equilibrium unit cell of the model under consideration. 
In other words, the pressure on the simulation box should be hydrostatic; the pressure tensor is diagonal
and each of the elements along the diagonal have the same value.
If this is not the case the system will suffer stresses  and the structure and thermodynamic
properties will not reach their equilibrium values.
This is achieved using the technique proposed by Parrinello and
Rahman \cite{PRL_1980_45_001196,JCP_1982_76_02662,SM_1983_17_1199} and extended to Monte Carlo by Yashonath and Rao \cite{MP_1985_54_0245}. 
Briefly, the shape of the simulation box is defined by a so-called $H$-matrix
representing the Cartesian coordinates of the vectors defining the simulation box.  
Each of the individual components of the $H$-matrix are adjusted
randomly, leading to changes in both the simulation box lengths and in the geometry.

\begin{figure}[ht]
\begin{center}
\includegraphics[height=400pt,width=500pt]{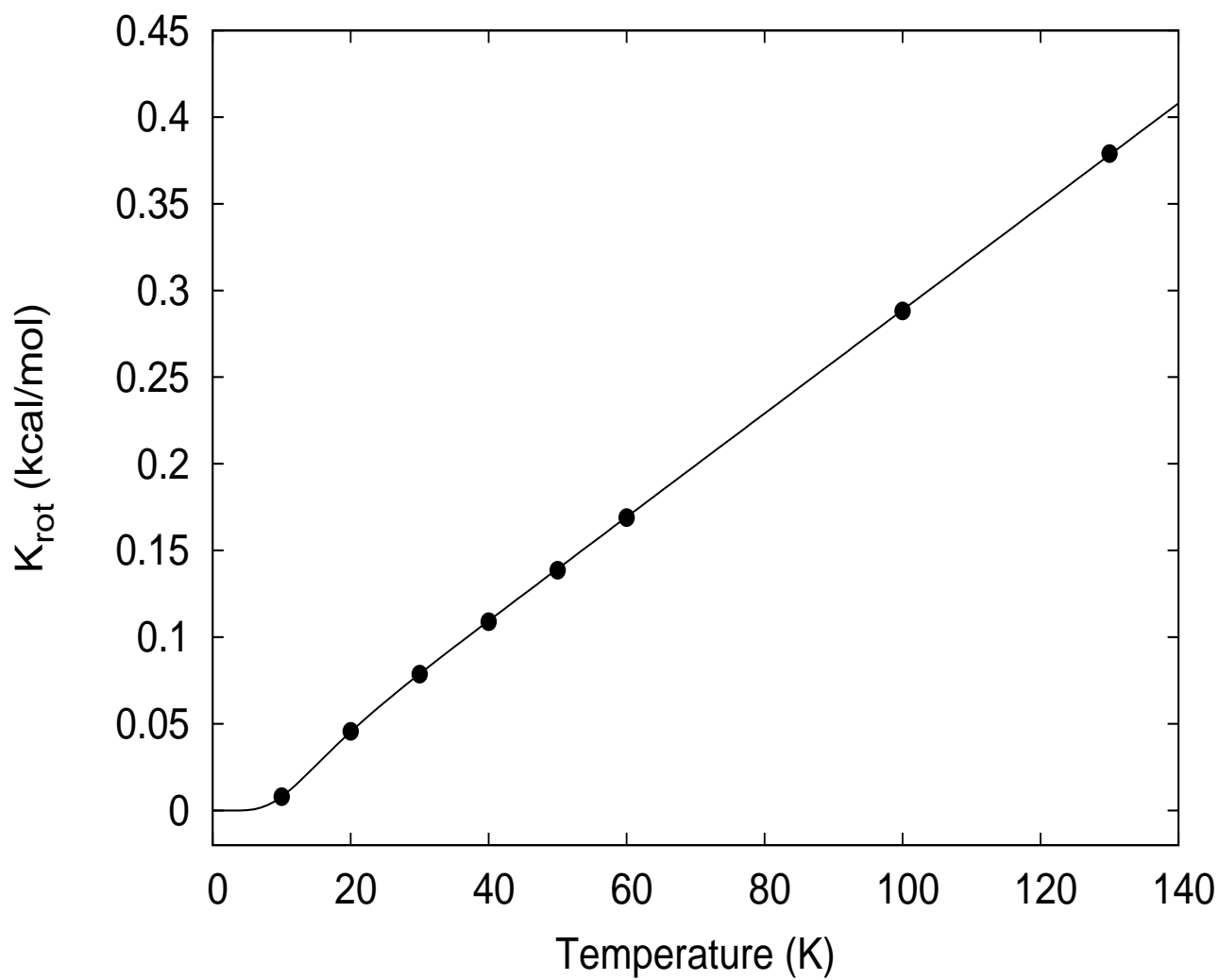}
\caption{\label{isolated_molecule} 
Kinetic rotational energy from PIMC simulations of the isolated H$_2$O
molecule (filled circles) as a function of temperature. Between 10 and 50 replicas (P) have been used, depending on the temperature.
There is good agreement between the simulation data and the rotational energy obtained from the theoretical
partition function of an asymmetric top having the  H$_2$O geometry (solid line).
The magnitude of  the error is less than the size of the symbols shown.
} 
\end{center}
\end{figure}

As a preliminary check that the M{\"u}ser and Berne propagator was implemented correctly 
the rotational energies were calculated for  an isolated H$_2$O molecule. 
In Fig. \ref{isolated_molecule} the rotational energies
computed from the exact expression of the
quantum partition function of an asymmetric top \cite{book_LevineMolecularSpectroscopy} 
(with the appropriate rotational constants) are compared
to those obtained from PIMC simulations.
As can be seen the agreement is excellent.
It should be noted that the present calculations do not include exchange
effects. However, these  are only expected to be relevant at temperatures below those that we have studied in this
work.
\section{Results}
\label{results}
A single state point has been simulated for each of the solid phases of water with the exception of 
ice X,  which cannot be described by a rigid model  \cite{N_1998_392_00258,N_1999_397_00503}. 
The results of these simulations are presented in Table \ref{table_ice_phases_1}. 
By comparing the densities obtained from classical TIP4P/2005 simulations to path integral 
simulations of the TIP4P/2005 model, which henceforth we shall denote as  TIP4P/2005$_{\mathrm{(PI)}}$,
it is clear that the introduction of  atomic quantum delocalisation effects reduces the density 
of the solid phase by about 
0.02 g/cm$^{3}$ for temperatures above 200K, and by $\approx0.03-0.04$g/cm$^{3}$ for temperatures in the 
range 75-170K.
Not surprisingly, quantum effects become increasingly evident as the temperature is reduced.
The various contributions to the total energy,  $E$, are also tabulated. 
As far as the translational kinetic energy component, $K_{\mathrm {translational}}$,  is concerned 
one can observe an increase of about 10\% for TIP4P/2005$_{\mathrm{(PI)}}$ with respect to 
TIP4P/2005 (i.e. $(3/2)RT$) at temperatures above 225K. As the temperature is lowered, this difference becomes
100\%. This is approximately true for all of the ices.
From these results one can conclude that the translational contribution to the heat 
capacity in quantum simulations is significantly less than the corresponding contribution in 
classical simulations.
If one looks at  the rotational kinetic energy contribution, 
$K_{\mathrm {rotational}}$, the differences are exaggerated even further;
ranging from about 100\% for `high' temperature ices, and increasing
to 600\% at low temperatures.
From this it is clear that the quantum contributions  are manifestly rotational 
in their nature, whilst translational effects are secondary in the solid phase.
Within a perturbative treatment the quantum contribution to the Helmholtz energy function
is related to the average of the forces divided by the masses for the translational contribution, 
and to the average of the torques divided by the principal moments of inertia for the orientational contribution \cite{AllenTildesleyChp10}. 
The mass of water is almost the same as that of neon, however, the quantum effects are  far more pronounced 
in water for the temperature range considered in this work \cite{JCP_2008_129_204502}.
The overwhelming reason for this difference is the strength and directionality of the hydrogen bond.
This, as well as the fact that the moments of the inertia tensor are quite small 
due to hydrogen having a very low mass.
The temperature dependence of the kinetic rotational energy is rather weak, so its 
contribution to the heat capacity is expected to be small. On the other hand 
the quantum contributions  to the potential energy are of the order of  1 kcal/mol at high temperatures,
which increases to  1.5 kcal/mol at low temperatures. Thus there is a significant
difference in  $E$ between the  TIP4P/2005 and the  TIP4P/2005$_{\mathrm{(PI)}}$ results,
amounting to  about 3 kcal/mol at low temperatures; half of which being due potential energy, and the other half kinetic. 

\begin{figure}[ht]
  \begin{center}
   \includegraphics[height=400pt,width=500pt]{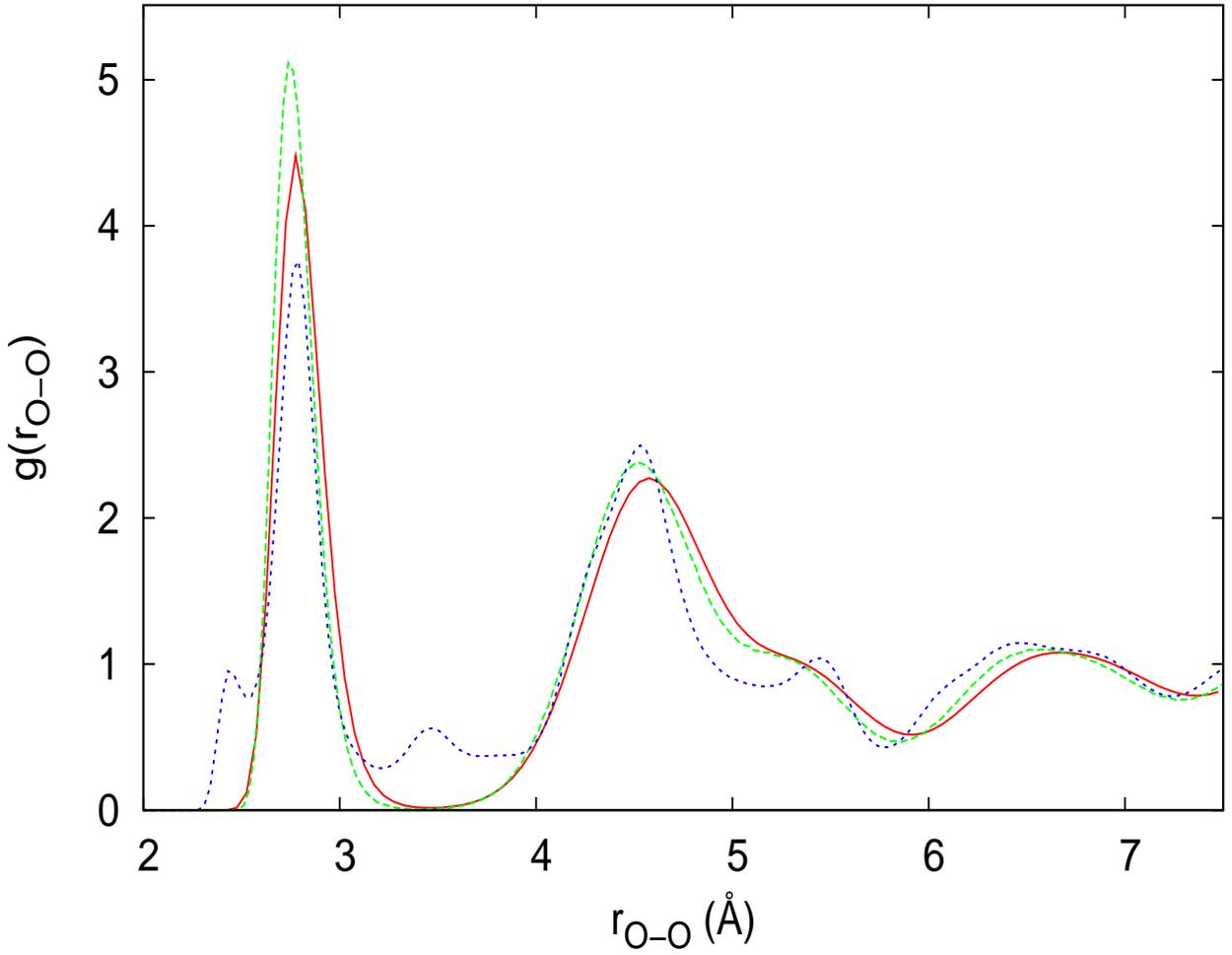}
   \caption{Radial distribution function of ice I$_h$ for TIP4P/2005 (dashed green line) and TIP4P/2005$_{\mathrm{(PI)}}$ (solid red line) at 250 K and p=0 bar. The blue dotted line 
corresponds to the experimental data of Soper at 220K \cite{CP_2000_258_0121}.}
   \label{fig:rdf_Ih}
  \end{center}
\end{figure}

\begin{figure}[ht]
  \begin{center}
   \includegraphics[height=400pt,width=500pt]{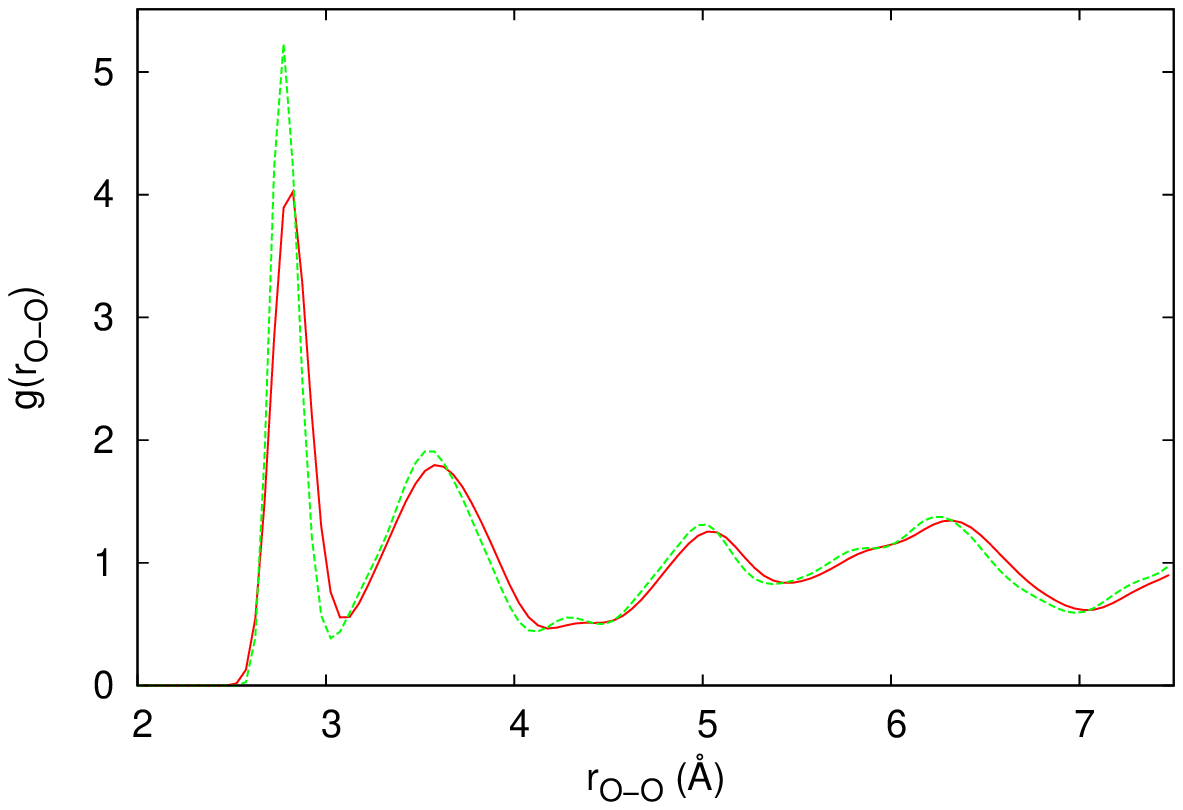}
   \caption{Radial distribution function of ice II for TIP4P/2005  (dashed green line) and TIP4P/2005$_{\mathrm{(PI)}}$ (solid red line) at 123 K and p=0 bar.}
   \label{fig:rdf_II}
  \end{center}
\end{figure}

\begin{figure}[ht]
  \begin{center}
   \includegraphics[height=400pt,width=500pt]{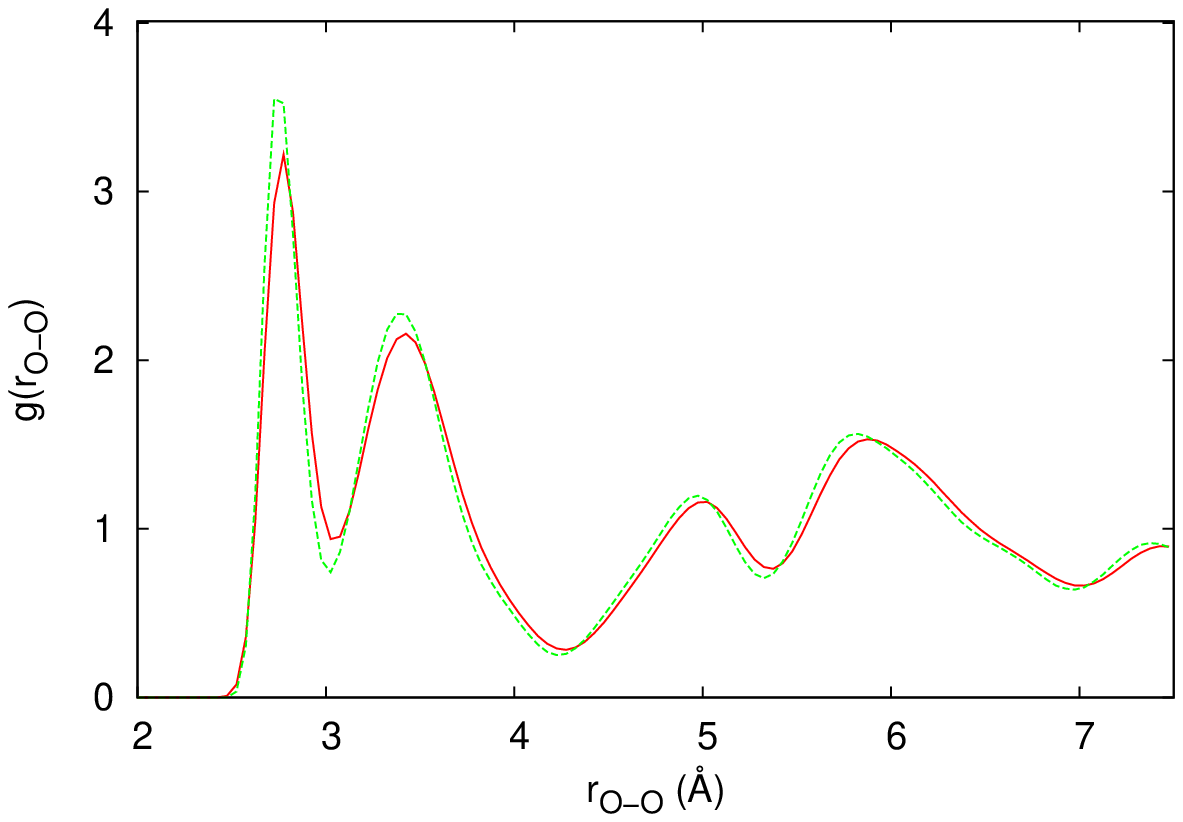}
   \caption{Radial distribution function of ice VI for TIP4P/2005 (dashed green line) and TIP4P/2005$_{\mathrm{(PI)}}$ (solid red line) at 225 K and p=11 kbar.}
   \label{fig:rdf_VI}
  \end{center}
\end{figure}

We shall now turn to the radial distribution functions. These histograms provide insights into 
the structure of a fluid on a 
molecular scale \cite{JCP_1984_81_04087,PCCP_2005_07_1450}.
One of the first simulation studies of such functions for water using path integral simulations 
was undertaken by Kuharsky, Rossky 
and co-workers \cite{CPL_1984_103_0357,JCP_1985_82_05164,JCP_1991_95_03728,JCP_1985_82_05289} for the ST2 model.
Given the low scattering factor of hydrogen, the oxygen-oxygen ($g_{\rm OO}$) is the distribution function
most accessible experimentally.
Here we present the oxygen-oxygen radial distribution function 
for ices I$_h$, II and VI (Figures \ref{fig:rdf_Ih}-\ref{fig:rdf_VI})
for classical TIP4P/2005 and TIP4P/2005$_{\mathrm{(PI)}}$. 
For ice  I$_h$ the experimental radial distribution function has also been plotted,
using the data provided by Soper  \cite{CP_2000_258_0121} at 220K. 
To the best of our knowledge as yet there are no experimental radial distribution functions
available in the literature for ices II and VI.
In Table \ref{table_OOrdf} details are given for specific points located 
along the oxygen-oxygen radial distribution
function curves for ice I$_h$.
On going from classical simulations to path integral simulations the location of the 
first two peaks shifts to slightly larger distances.
Furthermore, there is a notable reduction in the height of  these peaks when 
quantum contributions are incorporated.
Similar findings have been published previously for water and for simulations of 
TIP4P$_{\mathrm{(PI)}}$ of ice I$_h$ by Hern\'{a}ndez de la Pe\~{n}a et al. \cite{JCP_2005_123_144506}. 
This softening of the distribution functions goes hand-in-hand with the reduction
in the density of the ices in the PIMC calculations.
It is interesting to speculate whether the addition of the
small (and somewhat unusual) first peak in the ice I$_h$ experimental data with the much larger second peak
would place the simulation results in a more favourable light.

\begin{figure}[ht]
  \begin{center}
   \includegraphics[height=400pt,width=500pt]{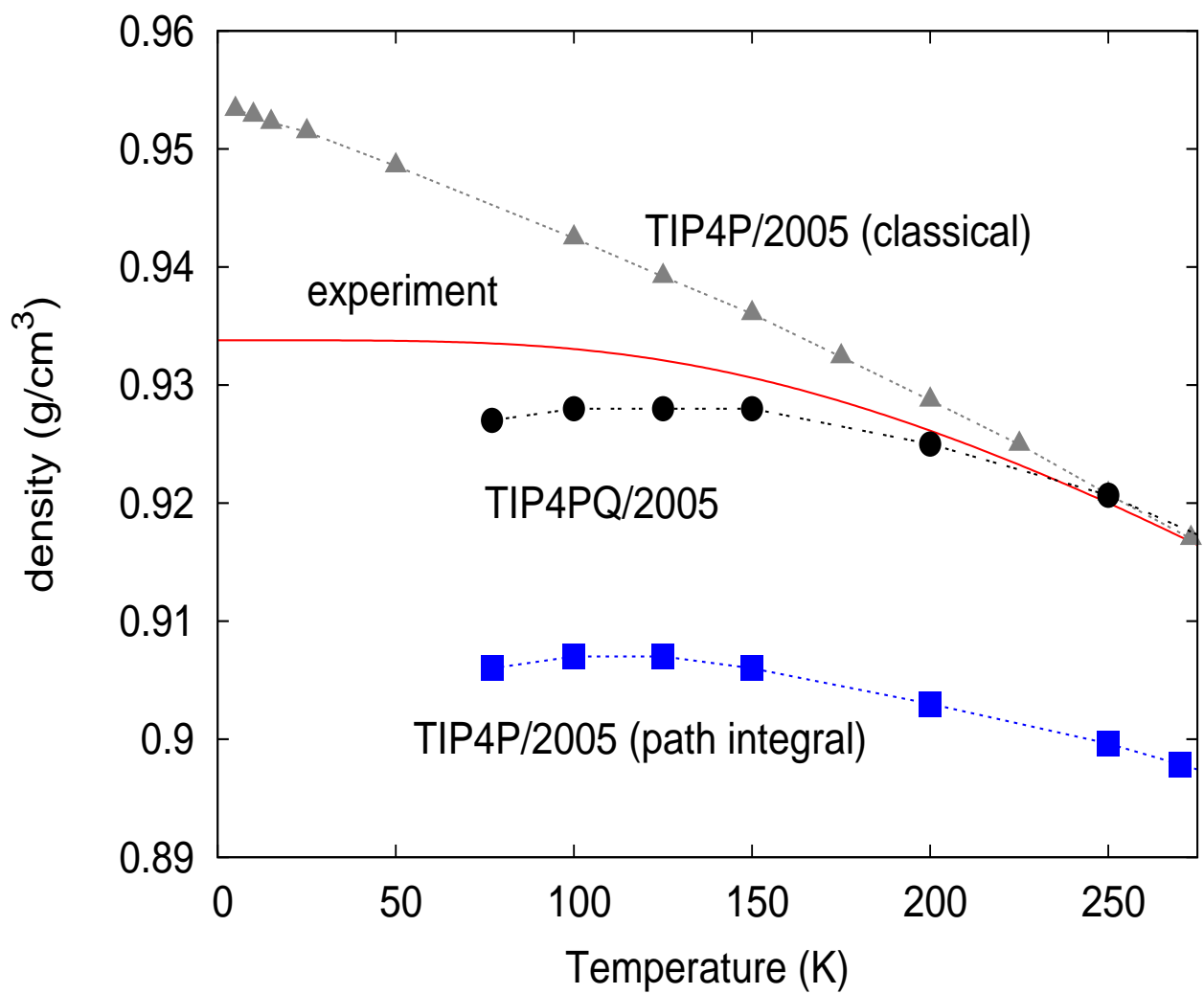}
   \caption{Equations of state for ice I$_h$ at $p=1$ bar. Classical TIP4P/2005 model (grey dot-dashed line / filled triangles) \cite{JPCC_2007_111_15877}, experimental data (red solid line) \cite{JPCRD_2006_35_1021}, TIP4P/2005$_{\mathrm{(PI)}}$ (blue dotted line/ filled squares) and the new TIP4PQ/2005 model (black double-dotted line / filled circles).
The error in the density is of order $\pm 0.002$  g$\cdot$cm$^{-3}$.}
   \label{fig:ice_eos}
  \end{center}
\end{figure}

A consequence of the third law of thermodynamics is that the coefficient of 
thermal expansion, $\alpha$, tends to zero when the temperature goes to zero. 
Experimentally one finds that there is very little variation in the density of 
ice I$_h$ in the temperature range 0-125K. Classical simulations are unable to 
capture this, as can be seen in the low temperature equations of state published in Ref \onlinecite{JPCC_2007_111_15877},
where the density of ice continues to increase as the temperature is lowered.
Here we have performed simulations of  TIP4P/2005$_{\mathrm{(PI)}}$ for temperatures in the 
range 77-200K along the atmospheric pressure isobar for a number of ices. These results are presented 
in Table \ref{table_zero}.   
In particular, the equation of state of ice I$_h$ is plotted in Fig. \ref{fig:ice_eos}
along with classical  \cite{JPCC_2007_111_15877} and experimental results \cite{JPCRD_2006_35_1021}.
One can see a dramatic reduction in the density between classical  TIP4P/2005 and TIP4P/2005$_{\mathrm{(PI)}}$
simulations. However, the most important difference is that the density is almost independent 
of the temperature below $\approx 125$K, in other words,  $\alpha$ tends to zero.
Given the fact that the TIP4P/2005 model was parameterised for classical simulations, it is 
no surprise that the  TIP4P/2005$_{\mathrm{(PI)}}$ results show a significant deviation from the  experimental values.
That said, the   TIP4P/2005$_{\mathrm{(PI)}}$ curve is, more or less, parallel to the experimental
curve, strongly suggesting that a re-parameterisation of the TIP4P/2005 model could improve
these results by shifting the TIP4P/2005$_{\mathrm{(PI)}}$ curve to higher  densities.
It is worth mentioning that the 100K state point for the TIP4P/2005$_{\mathrm{(PI)}}$ 
model seems to be slightly more dense than the 77K state point. It has been suggested that 
there is a temperature of maximum density in the ice phase \cite{JCP_1998_108_04887,JCP_2004_121_07926}, however, 
longer and more detailed simulations would have to be undertaken to establish whether our
results do indeed reflect this or not,
given that this curvature could well be due to the statistical uncertainties in the simulation results.

\begin{figure}[ht]
  \begin{center}
   \includegraphics[height=400pt,width=500pt]{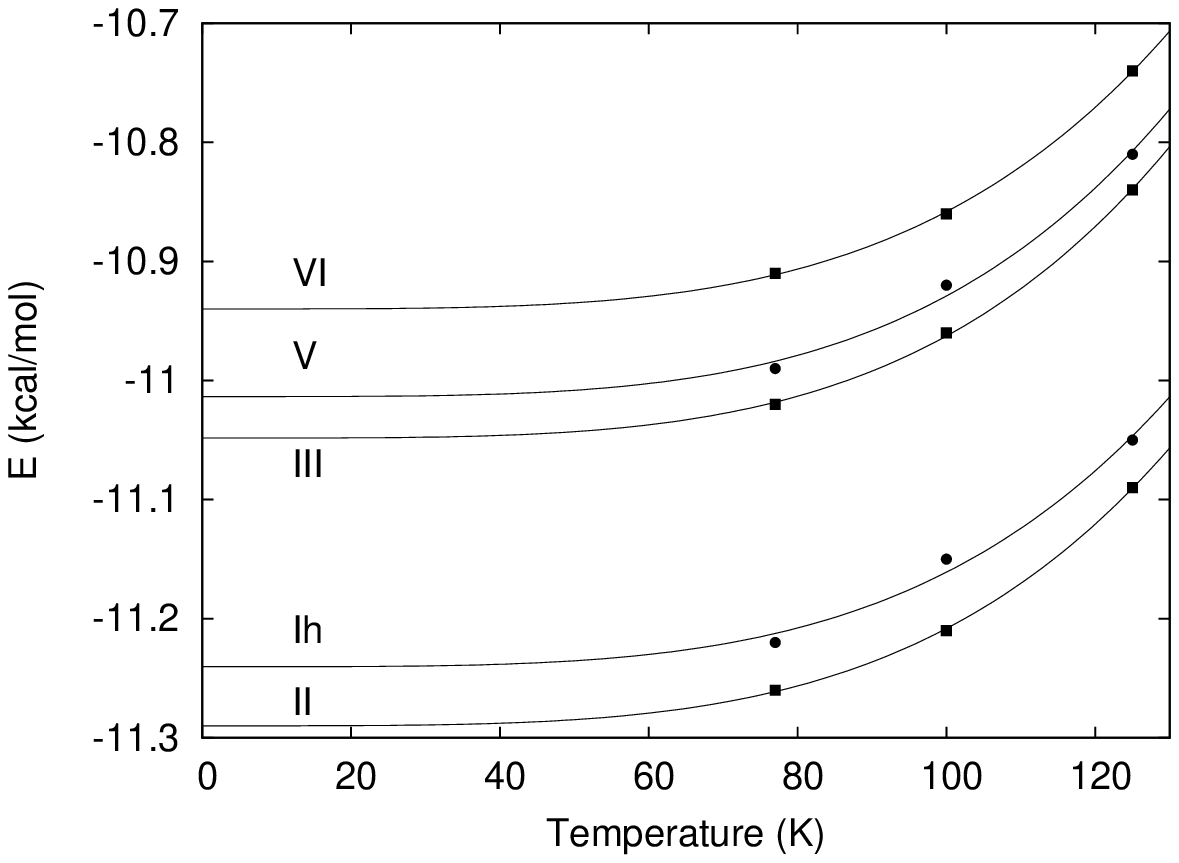}
   \caption{Plot of the total energy of ices I$_h$, II, III, V and VI at low temperatures for p=1 bar for TIP4P/2005$_{\mathrm{(PI)}}$.
Lines correspond to the fit $E=a+bT^4$.
The error in the total energy is of order $\pm 0.04$  kcal/mol.}
   \label{fig:zero:2005}
  \end{center}
\end{figure}

In 1984 Whalley estimated the thermodynamic properties of ices at 0K.
This estimate was made after analysing the experimental coexistence
curves between ices at low temperatures  \cite{JCP_1984_81_04087} and realising 
that at 0K phase transitions occur with zero enthalpy change. 
By assuming that the volume and internal energy difference between 
ices is largely unaffected by pressure (a quite reasonable approximation) Whalley
was able to estimate the energies and densities of ices at 0K and zero pressure. 
Such a calculation is useful as it allows one to obtain an idea of the form
of the phase diagram at low temperatures by examining the relative stability 
of the ice phases.
Thus one can estimate the coexistence pressure between two ice phases at zero kelvin using the approximation
\begin{equation}
\label{zero_pressure_eq}
p_{\mathrm{eq}} = \left. \frac{-\Delta U}{\Delta V} \right\vert_{p=0} 
\end{equation}
More recently a similar analysis was undertaken \cite{JCP_2007_127_154518}
for a number of popular empirical models of water.
For the SPC/E and  TIP5P models ice II was found to be more stable than ice I$_h$,
however, for TIP4P/2005 ice  I$_h$, as is the experimental situation, was more stable than ice II.
Here simulations were performed at 125, 100 and 77K for TIP4P/2005$_{\mathrm{(PI)}}$ 
(for technical reasons PIMC simulations at 0K are infeasible, given the number of 
beads required).
Assuming that the heat capacity, $C_{p}$, follows the Debye law, i.e $C_{p} \propto T^{3}$,  
then it follows that the enthalpy  should scale as   $T^{4}$. 
Note that the internal energy and enthalpies are almost indistinguishable at room pressure; the 
$pV$ term is negligible compared to the internal energy term. 
In Fig. \ref{fig:zero:2005} the internal energies from  Table \ref{table_zero} are plotted as a function of the temperature
for  TIP4P/2005$_{\mathrm{(PI)}}$ and the estimated values at 
0K, obtained from a fit of the form $E=a+bT^4$,  are given in  Table \ref{zero_comparison}. The relative energies between ices obtained at 0K
from the extrapolation procedure described above are quite similar to those obtained from
the simulations results at 77K. 
The inclusion of quantum effects consistently increases the energy at 0K of the ice phases by $\approx  3.5$ kcal/mol.
However, for ices II, III, V and VI the relative energy remains largely unchanged; differing by only $\approx$ 0.1 kcal/mol from the classical values. 
The zero point energies of ices II, III, V and VI are quite similar
and are expected to have very little effect on the 
relative stability of the ice phases.
This is not the case for ice I$_h$,  atomic quantum delocalisation effects destabilise ice I$_h$ with respect to ice II, the difference now being  $\approx$ 0.26 kcal/mol.
For example,  for TIP4P/2005$_{\mathrm{(PI)}}$  ice II replaces  I$_h$ as the most stable ice phase at low 
temperatures.
Given the fact that quantum effects stabilise ice II with respect to ice I$_h$ implies that 
for the TIP3P, SPC/E, and TIP5P models  the inclusion of  atomic quantum delocalisation effects would further deteriorate
their phase diagrams; the ice I$_h$ phase being stable only for large negative pressures 
and ice II dominating the low temperate atmospheric pressure isobar.
An interesting question is precisely why ice I$_h$ is more affected than the rest of the ices by  these
 atomic quantum delocalisation effects. As discussed previously, within a perturbative treatment
the effect of  atomic quantum delocalisation effects can be expressed as the average of forces and 
torques on the molecules divided by their masses or principal moments of inertia. Since the 
mass and inertia tensors are the same, regardless of the ice phase considered, differences 
between ices must be due to differences in forces and torques between molecules. 
In all the ices each water molecule forms four hydrogen 
bonds with its nearest neighbours. For ice I$_h$, the four nearest neighbours form an
almost perfect tetrahedron. However, for ices II, III, V and VI, the four nearest bonds 
form a distorted tetrahedron \cite{url_Chapman_Water}, resulting in weaker hydrogen bonds (even though 
they are more dense than ice I$_h$).
It is the strength of the I$_h$ hydrogen bonding that is showing up in the 
quantum contributions.

The results presented thus far have elucidated the quantum contributions to the 
properties of the solid phases of water.
The TIP4P/2005 model used in this study was originally parameterised to reproduce as faithfully as possible 
the experimental results for water using classical simulations. Thus in some implicit way, 
quantum contributions form part of the make-up of this model.
It is no surprise that an explicit introduction of 
quantum effects will degrade the qualitative aspects of this model, which is 
exactly what we have seen in this work using  TIP4P/2005$_{\mathrm{(PI)}}$.
We have witnessed that quantum effects decrease both the structure and the density of the ices 
as the temperature is lowered, and that they modify the relative stability of ices I$_h$ and II.
Originally the TIP4P/2005 model was created by examining the derivatives of the parameters of the model
for a number of properties, and then, via a least squares fit, the optimum values for the parameters
are obtained. These properties include the density and the coexistence lines obtained from values of the 
Helmholtz energy function. However, here we do not yet have access to the coexistence lines for the 
TIP4P/2005$_{\mathrm{(PI)}}$ model so in developing the new TIP4PQ/2005 model
a modest, and quite  probably sub-optimal, change in the parameters was called for.  

There is a veritable plethora of classical empirical models for water in the literature.
In contrast, there is a paucity of quantum empirical models. It is worth making a mention
of three of these quantum models;  a re-parameterisation of
a flexible version of the SPC/Fw model \cite{JCP_2006_125_184507}, the second is a re-parameterisation
of the rigid TIP5P model \cite{JCP_2001_115_10758}, and the third is a series of flexible and polarisable potential
models named  TTM2-F \cite{JCP_2002_116_05115} and TTM3-F \cite{JCP_2008_128_074506},  obtained from  fits to the potential energies of water clusters obtained
from first principle calculations. For both the SPC and the TIP5P re-parameterisations
the essential difference was that the  dipole moment
of the molecule was increased, whilst maintaining the remaining parameters of the potential constant.
The basic idea is that since  atomic quantum delocalisation effects reduce the density and internal energy
of the system, increasing the charge is a simple way of `re-compensating' for these changes, 
coaxing the model back to being its former  self.
It was with this in mind that the TIP4PQ/2005 model was created.
The only difference between the TIP4P/2005 and the TIP4PQ/2005 models is in the 
dipole moment (see Table \ref{tip4p_2005_parameters}),
which  was increased from 
2.305D to 2.38D. This was achieved by a 0.02$e$ increase in the charge of the protons.
Similar increases in the dipole moments of water (of about 0.08-0.10D)
were used in the aforementioned quantum versions of SPC   \cite{JCP_2006_125_184507} and TIP5P models \cite{JCP_2001_115_10758}.
Such an increase in the charge may not be necessary in a flexible model where, as stated by Mahoney and Jorgensen,
``...although quantum effects make the density behaviour of the rigid model worse, 
they improve the density behaviour of the flexible model." \cite{JCP_2001_115_10758}.
This interplay between an increase in the  dipole moment and flexibility has also been commented upon by other authors \cite{CPL_1996_250_0019,JML_2002_101_0219}.
Obviously this new model is only suitable for quantum simulations of water.

In Table \ref{table_ice_phases_2} the state points for the ice phases are recalculated using this new TIP4PQ/2005
model. When compared to the experimental values 
\cite{ACSB_1994_50_644,
JdP_1987_48_C1-631,
JAC_2005_38_0612,
JCP_1993_98_04878,
JCP_1981_75_05887,
JCP_1990_92_01909,
JCP_1984_81_03612,
N_1987_330_00737_nolotengo,
JCP_1996_104_10008,
N_1998_391_00268,
S_2006_311_01758}
the results are really quite good over the whole range of temperatures and pressures.
The average quadratic deviation between experimental and predicted densities (excluding ice VII) is 0.01  g/cm$^{3}$ for the 
classical TIP4P/2005 model, which becomes 0.03  g/cm$^{3}$   for the  TIP4P/2005$_{\mathrm{(PI)}}$ model.
For the re-parameterised TIP4PQ/2005 model the quadratic deviation is once again 0.01  g/cm$^{3}$, recovering
the situation for the classical model for the state points considered.   
In Table \ref{unit_cell} the unit cell parameters for the TIP4PQ/2005 model for a selection of ice phases 
have been provided and are also seen to be rather good when compared to the experimental values.

\begin{figure}[ht]
  \begin{center}
   \includegraphics[height=400pt,width=500pt]{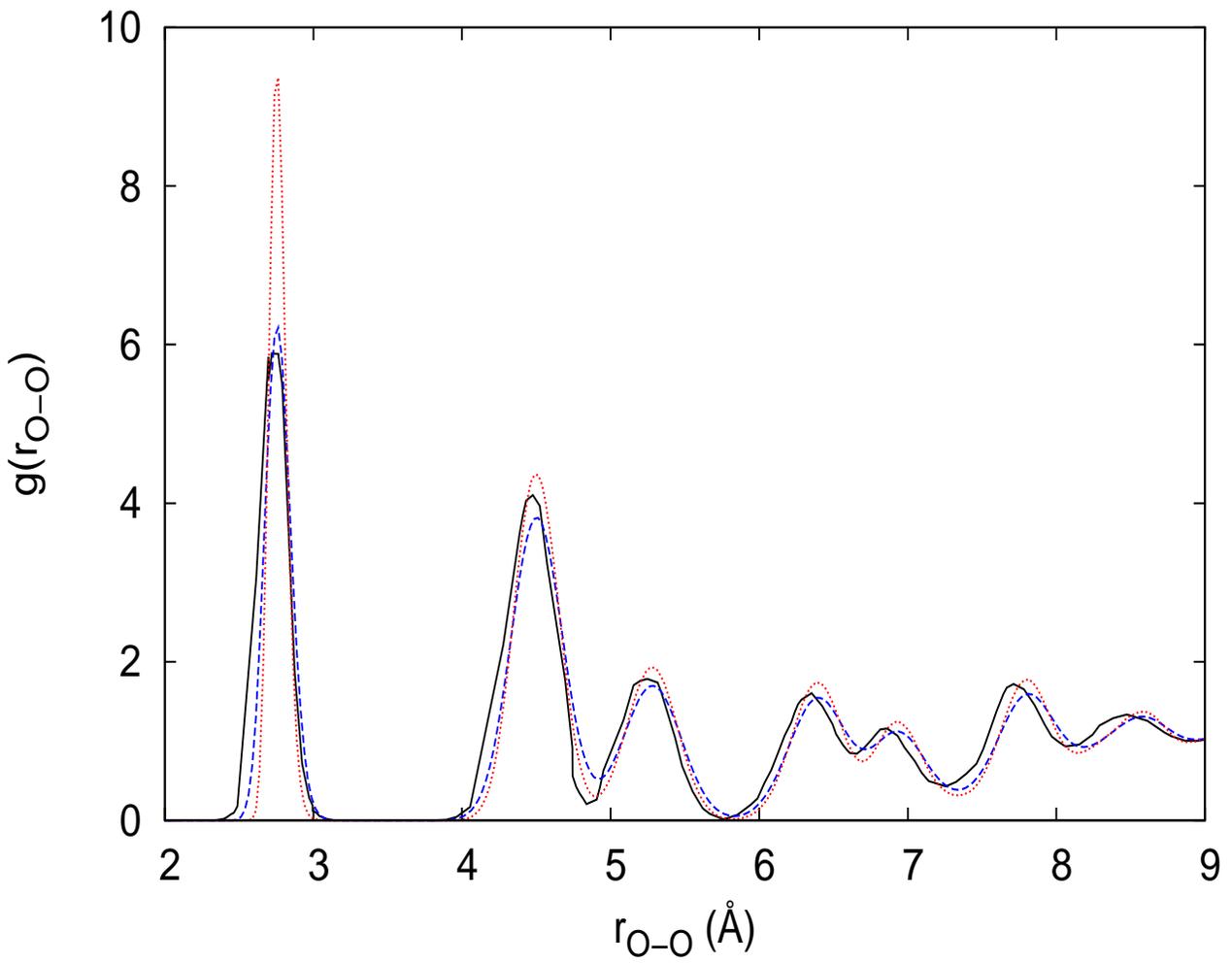}
   \caption{Radial distribution function of ice I$_h$ for the TIP4PQ/2005 model using PIMC (dashed blue line) compared with the classical TIP4P/2005 model (dotted red line) and with experimental data  (solid red line) \cite{JCP_1976_64_01106}
at 77K and p=1 bar.}
   \label{fig:rdf_Ih_Q_Narten}
  \end{center}
\end{figure}

In Fig  \ref{fig:ice_eos} the equation of state for ice I$_h$ is plotted. The TIP4PQ/2005 state points are equidistant
from those of  TIP4P/2005$_{\mathrm{(PI)}}$, but they are now much closer to the experimental values, 
with a deviation of around  0.005 g/cm$^{3}$,
which amounts to a difference of only 0.8\% with respect to the experimental value. 
Given the curvature of the equation of state, in line with the 
third law of thermodynamics, and the small difference between the TIP4PQ/2005 densities and the experimental 
results, leads us to believe that this is one of the best theoretical descriptions of ice I$_h$ thus far seen in the literature.
This is not to say that in the future this cannot be improved upon, for example via the inclusion of 
flexibility, polarisability etc. in the molecular model. 
In Fig. \ref{fig:rdf_Ih_Q_Narten} the oxygen-oxygen radial distribution function of ice I$_h$  at 77K is compared to the 
experimental results of Narten \cite{JCP_1976_64_01106}, and the results are acceptable
almost all the way up to 9{\AA}. The most notable difference can be seen in the  height of  the 
first peak; which drops from 9.37 for classical TIP4P/2005, down to 6.21 for TIP4PQ/2005, compared to 5.95 experimentally  \cite{JCP_1976_64_01106}.

\begin{figure}[ht]
  \begin{center}
   \includegraphics[height=400pt,width=500pt]{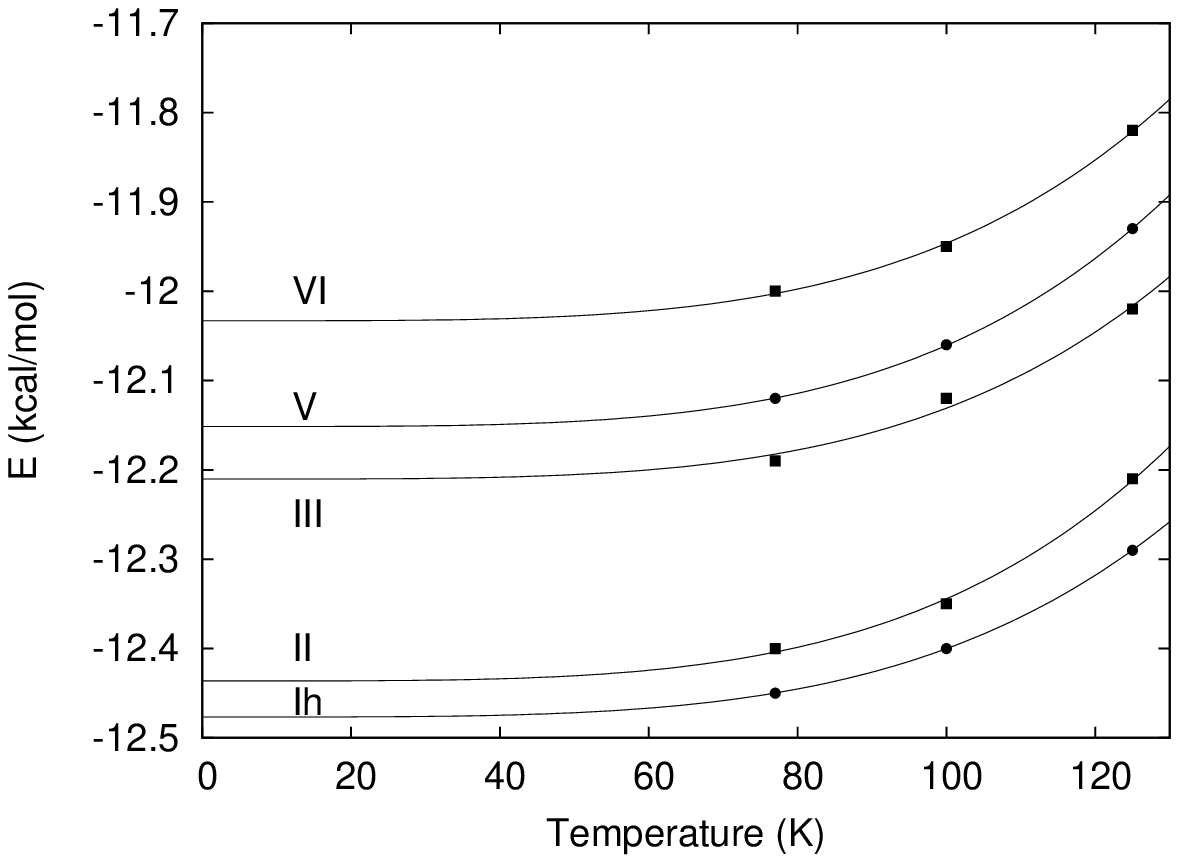}
   \caption{Plot of the total energy of ices I$_h$, II, III, V and VI at low temperatures for p=1 bar for the TIP4PQ/2005 model.
Lines correspond to the fit $E=a+bT^4$.
The error in the total energy is of order $\pm 0.04$  kcal/mol.}
   \label{fig:zero_Q}
  \end{center}
\end{figure}

In an analogous study to that for 0K for  TIP4P/2005$_{\mathrm{(PI)}}$ the 
relative stability of ices  I$_h$, II, III, V and VI  at low temperatures has been tabulated in Tables  \ref{zero_comparison} and  \ref{table_zero_2}
and plotted in Fig. \ref{fig:zero_Q}. 
As can be seen the relative energy between ice II and the remainder of the ices is similar to 
that of TIP4P/2005$_{\mathrm{(PI)}}$.
The most significant result is that for TIP4PQ/2005 ice I$_h$ regains
its rightful place as the most stable ice phase. Experimentally the energy difference between 
I$_h$ and II is 0.014 kcal/mol, which for TIP4PQ/2005 becomes 0.04 kcal/mol.
In Table \ref{table_pressure_zero_2} results for the 0K coexistence pressures, calculated using equation \ref{zero_pressure_eq},   are presented.
It can be seen that both the energies (Table \ref{zero_comparison}) and the coexistence pressures (Table \ref{table_pressure_zero_2}) for various transitions 
are substantially better than the values provided by classical simulations of the TIP4P/2005 model,
in particular for the I$_h$-II transition.
This gives us confidence that the TIP4PQ/2005 could well produce a respectable
global phase diagram in the future.

\section{Conclusions}
\label{conclusions}

This work addresses a series of physical properties of water that vary with the inclusion of  atomic quantum delocalisation effects, 
which were introduced to the TIP4P/2005 model using path integral Monte Carlo simulations.
Quantum simulations have been undertaken for 
all of the ice phases of water, with the exception of ice X, 
for the TIP4P/2005 model, and for the new TIP4PQ/2005 model.
Using the  M{\"u}ser and Berne propagator for rigid asymmetric tops, 
various properties of these ices have been examined.

It has been found that the radial distribution functions become
more `washed-out' when quantum effects are taken into account.
In other words, the peaks become lower and wider
and shift to slightly larger distances.
This goes hand-in-hand with a reduction in density for the 
quantum solid; by  $\approx 0.02$ g/cm$^3$  for temperatures above 150K, 
and $\approx$ 0.04 g/cm$^3$ below 100K.

If a classical empirical model is tailored to reproduce the 
experimental ice densities at a temperature close to the melting point, as the temperature is 
reduced the model will start to fail 
(such is the case, for example, of the TIP4P/2005 model \cite{JPCC_2007_111_15877}). 
This is due to the fact that classical simulations are 
unable to satisfy one of the
consequences of the third law of thermodynamics, 
namely that the coefficient of thermal
expansion, $\alpha$, tends to zero as the temperatures approaches zero Kelvin.
It can be seen that the PIMC simulations now, to a good degree, 
correctly describe the low temperature physics of this model.

The translational component of the kinetic energy bears a passing resemblance 
to the classical value of $(3/2)RT$, whereas the rotational component 
is markedly larger.

There is a particularly pronounced effect in the relative stabilities of
ices I$_h$ and II, where the stability of ice II is enhanced by the inclusion 
of  atomic quantum delocalisation effects.

In this work a re-parameterisation of the TIP4P/2005 model is provided
that `compensates' for the quantum effects so as to maintain the 
quantitative performance of the TIP4P/2005 model, whilst at  the same time
reproducing the correct physics at low temperatures.
In this new model, which we have called TIP4PQ/2005, the only 
parameter to have changed is that of the dipole moment; the charge on the hydrogen atom 
has been increased by 0.02$e$, thus the dipole moment increases from  2.30D
to 2.38D.

In this paper it is shown that the TIP4PQ/2005 model provides a good description
of the densities  of the ice phases for the state points considered.
The ice I$_h$  $p=1$ bar isobar has been calculated and the tendency for $\alpha$ to become
zero is now present in the equation of state.
This new model also correctly describes the relative stabilities of  ices  I$_h$  and II.
An extrapolation indicates that at 0K  I$_h$ is more stable than ice II by  0.04 kcal/mol (compared to 0.014 kcal/mol experimentally).
The inclusion of quantum effects substantially improves the overall description of all of the ice phases
studied here. The TIP4P/2005 does a reasonable job, but the TIP4PQ/2005 is clearly superior.
   This paper can be regarded as a first step in introducing atomic quantum delocalisation effects
   in the description of the solid phases of water. 
   However, it is by no means the last word, since obviously water is a flexible molecule. 
   In our opinion the results in the present manuscript could be very useful as a 
   point of departure for the development of a flexible model of water for use in 
   path integral simulations,
   and provides valuable material from which to make comparisons. Such comparison would 
   establish just how much of the quantum effects in water are due 
   to intra and how much is due to the  intermolecular degrees of freedom.

\section{Acknowledgements}
\label{acknowledgements}
The authors would like to thank J. L. Abascal  for insightful conversations and M. I. J. Probert  for his hospitality and illuminating
discussions whilst one of the authors, C. V. was in York.
We would also like to thank the reviewers for their useful comments regarding the manuscript.
This work has been funded by grants FIS2007-66079-C02-01
and FIS2006-12117-C04-03 from the DGI (Spain), S-0505/ESP/0299 (MOSSNOHO) from the Comunidad Autonoma de Madrid,
and 910570 from the Universidad Complutense de Madrid. E. G. N. would like to thank the MEC 
for a Juan de la Cierva fellowship.

\clearpage
\begin{table}
\begin{center}
\caption{Parameters for both the TIP4P/2005 and the TIP4PQ/2005 models. The distance between the oxygen and hydrogen sites
is $d_{\mathrm{OH}}$. The angle, in degrees,  formed by hydrogen, oxygen, and the other hydrogen atom is
denoted by $\angle$H-O-H. The Lennard-Jones site is located on the oxygen with parameters $\sigma$ and $\epsilon$.
The charge on the proton is $q_{\mathrm{H}}$. The negative charge is placed in a point M at a distance
$d_{\mathrm{OM}}$ from the oxygen along the H-O-H bisector.}
\label{tip4p_2005_parameters}
\begin{tabular}{lllllll}
\hline
Model &$d_{\mathrm{OH}}$ (\AA) &\hspace{0.2cm} $\angle$H-O-H &\hspace{0.2cm} $\sigma$(\AA)&\hspace{0.2cm} $\epsilon/k_{B}$(K)& \hspace{0.2cm} $q_{\mathrm{H}}$(e)& \hspace{0.2cm}$d_{\mathrm{OM}}$(\AA) \\
\hline
TIP4P/2005 & 0.9572 & \hspace{0.2cm}104.52 & \hspace{0.2cm}3.1589 & \hspace{0.2cm}93.2 &\hspace{0.2cm} 0.5564 &\hspace{0.2cm} 0.1546\\
TIP4PQ/2005 & 0.9572 & \hspace{0.2cm}104.52 & \hspace{0.2cm}3.1589 & \hspace{0.2cm}93.2 &\hspace{0.2cm} 0.5764 &\hspace{0.2cm} 0.1546\\
\hline
\end{tabular}
\end{center}
\end{table}

\clearpage
\begin{sidewaystable}
\begin{center}
\caption{\label{table_ice_phases_1}Results for the TIP4P/2005$_{\mathrm{(PI)}}$ model for the systems studied, along with a comparison to classical results for the same model.
 All energies are in units of kcal/mol  and the densities are in g$\cdot$cm$^{-3}$. 
The errors (in kcal/mol) are ${\mathcal {O}}(0.003)$ in  $K_{\mathrm {translational}}$,  ${\mathcal {O}}(0.02)$ in  $K_{\mathrm {rotational}}$,  
${\mathcal {O}}(0.02)$ in  $U$,  ${\mathcal {O}}(0.04)$ in  $E$ and  ${\mathcal {O}}(0.002)$ g$\cdot$cm$^{-3}$ in  $\rho$ . 
}
\begin{tabular}{lllllllllllr}
\hline
Phase (\& N$^{\mathrm{o}}$ molecules) &  $T$ (K) & $p$ (bars) &   $(3/2)RT$  & $K_{\mathrm {translational}}$ &  $K_{\mathrm {rotational}}$   & $K_{\mathrm {total}}$  &  $U$  &  $E$  & $U$ (classical)    & $\rho$ (path-integral) & $\rho$ (classical) \\
\hline
I$_h$ (432) &  250 	&   0  		  	& 0.75 & 0.83  &1.39 & 2.22 & -12.38&-10.17&             -13.35&0.899 & 0.920   \\
I$_c$ (216) &  78  	&   0  		  	& 0.23 & 0.45  &1.36 & 1.81 & -13.03&-11.22&             -14.58&0.906 & 0.943 \\   
II (432)   &  123 	& 0   		   	& 0.37 & 0.51  &1.26 & 1.77 & -12.83&-11.06&             -14.07&1.160 & 1.198   \\
III (324)  &  250 	& 2800 		 	& 0.75 & 0.83  &1.35 & 2.18 & -12.15& -9.96&             -13.06&1.141 & 1.159  \\
IV  (432)  &  110 	& 0   		 	& 0.33 & 0.49  &1.25 & 1.74 & -12.44&-10.70&             -13.74&1.248 & 1.292 \\
V  (504)   &  237.65 & 5300 		 	& 0.71 & 0.80  &1.35 & 2.14 & -12.19&-10.04&             -13.21&1.240&  1.271\\
VI (360)   &  225 	& 11000  	  	& 0.67 & 0.78  &1.34 & 2.12 & -12.21&-10.10&             -13.11&1.356 & 1.379  \\  
VII (432)  &  300 	& 100000  		& 0.89 & 1.05  &1.44 & 2.49 & -9.32 &-6.83&               -9.95&1.767 & 1.782    \\
VIII (600) &  77  	& 24000  	 	& 0.23 & 0.49  &1.17 & 1.76 & -11.31&-9.65&              -12.50&1.573 & 1.616   \\
IX  (324)  & 165  	& 2800 		  	& 0.49 & 0.63  &1.33 & 1.96 & -12.80&-10.84&             -13.95&1.160 & 1.190  \\
XI  (360)  &  77   	& 0	    	 	& 0.23 & 0.45  &1.36 & 1.81 & -13.04&-11.23&             -14.60&0.906 & 0.945   \\
XII (540)  & 260  	& 5000  		& 0.77 & 0.86  &1.34 & 2.20 & -11.97&-9.77 &             -12.85&1.267 & 1.296  \\
XIII (504) & 80  	& 1  			& 0.24 & 0.44  &1.25 & 1.69 & -12.76&-11.07&             -14.16&1.217 & 1.261  \\
XIV (540)  & 80  	& 1  		 	& 0.24 & 0.44  &1.27 & 1.71 & -12.80&-11.09&             -14.25&1.280 & 1.331 \\
\hline
\end{tabular}
\end{center}
\end{sidewaystable}

\clearpage
\begin{table}
\begin{center}
\caption{\label{table_OOrdf}Oxygen-oxygen radial distribution function of ice I$_h$ for various water models at 250K and p=0 bar.}
\begin{tabular}{l|ll|ll|l}
\hline
Model     &  \multicolumn{2}{l}{peak 1} &  \multicolumn{2}{l}{peak 2} &     Reference \\  \cline{2-5}
          &  $r$  &  height          &  $r$  &  height                           &           \\
\hline
TIP4P (classical)          &    2.725\AA   & 4.707 &   4.525\AA   & 2.279  &\cite{PCCP_2005_07_1450} \\
TIP4P (path integral)      &    2.7625\AA  & 4.167 &   4.5625\AA     & 2.122   &This work\\
TIP4P/2005 (classical)     &    2.7375\AA  & 5.113 &   4.5125\AA   & 2.382  & This work \\
TIP4P/2005$_{\mathrm{(PI)}}$&    2.7875\AA  & 4.481 &   4.5875\AA   & 2.270  & This work \\
TIP4PQ/2005 	   &    	2.7625\AA  & 4.725 &   4.5375\AA   & 2.405  & This work \\
\hline
\end{tabular}
\end{center}
\end{table}

\clearpage
\begin{table}
\begin{center}
\caption{\label{table_zero}Results for the TIP4P/2005$_{\mathrm{(PI)}}$ model for the low temperature ice phases at a pressure of  1 bar. The energies are in units of kcal/mol and the densities are in g$\cdot$cm$^{-3}$.
The errors (in kcal/mol) are ${\mathcal {O}}(0.003)$ in  $K_{\mathrm {translational}}$,  ${\mathcal {O}}(0.02)$ in  $K_{\mathrm {rotational}}$,  
${\mathcal {O}}(0.02)$ in  $U$,  ${\mathcal {O}}(0.04)$ in  $E$ and  ${\mathcal {O}}(0.002)$ g$\cdot$cm$^{-3}$ in  $\rho$ . }
\begin{tabular}{llllllll}
\hline
Phase & $T$ (K)  &  $K_{\mathrm {translational}}$ &  $K_{\mathrm {rotational}}$   & $K_{\mathrm {total}}$  &  $U$  &  $E$   & $\rho$ \\
\hline
I$_h$   &  200   &   0.70 & 1.36 & 2.06 & -12.62  &      -10.56  &      0.903  \\
I$_h$   &  150   &   0.58 & 1.35 & 1.93 & -12.84  &      -10.91  &      0.906  \\
I$_h$   &  125   &   0.53 & 1.35 & 1.87 & -12.92  &      -11.05  &      0.907  \\
I$_h$   &  100   &   0.48 & 1.35 & 1.83 & -12.99  &      -11.15  &      0.907  \\
I$_h$   &  77    &   0.45 & 1.36 & 1.80 & -13.02  &      -11.22  &      0.906  \\
\hline
II    &  200     &   0.69 & 1.28 & 1.96 & -12.54  &      -10.57  &      1.145  \\
II    &  150     &   0.57 & 1.25 & 1.82 & -12.77  &      -10.95  &      1.155  \\
II    &  125     &   0.51 & 1.24 & 1.75 & -12.85  &      -11.09  &      1.159  \\
II    &  100     &   0.47 & 1.24 & 1.71 & -12.92  &      -11.21  &      1.163  \\
II    &  77      &   0.43 & 1.26 & 1.84 & -12.94  &      -11.26  &      1.165  \\
\hline
III    &  200   &    0.70 & 1.32 & 2.02 & -12.35  &      -10.34  &      1.106  \\
III    &  150   &    0.58 & 1.29 & 1.87 & -12.58  &      -10.70  &      1.116  \\
III    &  125   &    0.53 & 1.30 & 1.82 & -12.66  &      -10.84  &      1.122  \\
III    &  100   &    0.48 & 1.30 & 1.77 & -12.74  &      -10.96  &      1.125  \\
III    &  77    &    0.44 & 1.30 & 1.74 & -12.77  &      -11.02  &      1.130  \\
\hline
V    &  200   &      0.69 & 1.30 & 1.99 & -12.28  &      -10.29  &      1.204  \\
V    &  150   &      0.57 & 1.28 & 1.85 & -12.51  &      -10.67  &      1.217  \\
V    &  125   &      0.52 & 1.28 & 1.78 & -12.60  &      -10.81  &      1.222  \\
V    &  100   &      0.47 & 1.28 & 1.74 & -12.67  &      -10.92  &      1.225  \\
V    &  77    &      0.44 & 1.28 & 1.72 & -12.70  &      -10.99  &      1.227  \\
\hline
VI    &  200   &     0.69 & 1.27 & 1.96 & -12.19  &      -10.22  &      1.282  \\
VI    &  150   &     0.58 & 1.25 & 1.83 & -12.41  &      -10.58  &      1.296  \\
VI    &  125   &     0.51 & 1.25 & 1.76 & -12.50  &      -10.74  &      1.302  \\
VI    &  100   &     0.46 & 1.25 & 1.71 & -12.57  &      -10.86  &      1.306  \\
VI    &  77    &     0.43 & 1.26 & 1.69 & -12.60  &      -10.91  &      1.309  \\
\hline
\end{tabular}
\end{center}
\end{table}

\clearpage
\begin{table}
\begin{center}
\caption{\label{zero_comparison} Comparison of the energies, $E$, at 0K for a selection of phases for both the TIP4P/2005$_{\mathrm{(PI)}}$ and the TIP4PQ/2005 models as well as results for the classical TIP4P/2005 model \cite{JCP_2007_127_154518}.
 The energies are in units of kcal/mol. 
The lowest energy (most stable phase) is shown in bold font.
The lower section provides the relative energies with respect to ice II.
}
\begin{tabular}{lcccc}
\hline
Ice      &  \multicolumn{2}{l}{$E$ (0K estimate)} \\  \cline{2-4}
               & TIP4P/2005 &   TIP4P/2005$_{\mathrm{(PI)}}$  &  TIP4PQ/2005 &  Experimental \cite{JCP_1984_81_04087}    \\
\hline
I$_h$  & {\bf -15.059 }      & -11.240  &   {\bf -12.477}   & {\bf    -11.315}\\
II     &      -14.847  & {\bf  -11.290}  &       -12.436           &  -11.301\\
III     &     -14.741  &       -11.048  &        -12.210            & -11.100 \\
V      &      -14.644  &       -11.013  &        -12.152            & -11.088\\
VI     &      -14.513   &      -10.939  &        -12.033           &  -10.928 \\
\hline
I$_h$  & {\bf -0.212 }  &        0.050  &    {\bf -0.041}   &     {\bf -0.014}\\
II     &           0        &  {\bf 0}   &             0                  & 0 \\
III    &       0.106          & 0.242  &           0.226              & 0.201 \\
V       &      0.203          & 0.277  &           0.285              & 0.213\\
VI      &      0.334          & 0.351  &           0.403             &  0.373 \\

\hline
\end{tabular}
\end{center}
\end{table}

\clearpage
\begin{sidewaystable}
\begin{center}
\caption{\label{table_ice_phases_2}PIMC results for the TIP4PQ/2005 model for the systems studied and their relation
to the experimental densities.  All energies are in units of kcal/mol and the densities are in g$\cdot$cm$^{-3}$.
The errors (in kcal/mol) are ${\mathcal {O}}(0.003)$ in  $K_{\mathrm {translational}}$,  ${\mathcal {O}}(0.02)$ in  $K_{\mathrm {rotational}}$,  
${\mathcal {O}}(0.02)$ in  $U$,  ${\mathcal {O}}(0.04)$ in  $E$ and  ${\mathcal {O}}(0.002)$ g$\cdot$cm$^{-3}$ in  $\rho$ . 
}
\begin{tabular}{lllllllllrrr}
\hline
Phase &  $T$ (K) & $p$ (bars)  &  $(3/2)RT$  & $K_{\mathrm {translational}}$ &  $K_{\mathrm {rotational}}$   & $K_{\mathrm {total}}$  &  $U$  &  $E$     & $\rho$ (path-integral) & $\rho$ (experimental) & Reference \\
\hline
I$_h$ &  250 &   0                             & 0.75    &   0.83 & 1.45 & 2.28 & -13.74&-11.46&0.921 & 0.920  & \cite{ACSB_1994_50_644}  \\
I$_c$ &  78  &   0                & 0.23 &    0.46 & 1.43 & 1.89 & -14.33&-12.44&0.925 & 0.931  & \cite{JdP_1987_48_C1-631}  \\
II   &  123 & 0                  &  0.37 &    0.52 & 1.32 & 1.85 & -14.06&-12.21&1.185 & 1.190  & \cite{JAC_2005_38_0612} \\
III &  250 & 2800                &  0.75 &    0.84 & 1.41 & 2.25 & -13.44&-11.18&1.159 & 1.165  &\cite{JCP_1993_98_04878} \\
IV  &  110 & 0                   &  0.33 &    0.50 & 1.32 & 1.82 & -13.63&-11.81&1.276  &1.272&\cite{JCP_1981_75_05887}\\
V &  237.65 & 5300               &  0.71 &    0.81 & 1.41 & 2.22 & -13.43&-11.21&1.266 & 1.271     &\cite{JCP_1990_92_01909} \\
VI  &  225 & 11000               &  0.67 &    0.79 & 1.39 & 2.18 & -13.41 &-11.23&1.377 & 1.373  & \cite{JCP_1984_81_03612}  \\
VII   &  300 & 100000             & 0.89 &    1.05 & 1.47 & 2.52 & -10.37&-7.85 &1.780 & 1.880  & \cite{N_1987_330_00737_nolotengo} \\
VIII  &  77  & 24000              & 0.23 &     0.50 & 1.23 & 1.73 & -12.28&-10.56&1.592 &  1.628 (at 10K) & \cite{JCP_1984_81_03612}\\
IX    & 165  & 2800               & 0.49 &    0.64 & 1.39 & 2.04 & -14.07&-12.03&1.182 & 1.194 & \cite{JCP_1993_98_04878}  \\
XI   &  77   & 0                  & 0.23 &    0.46 & 1.43 & 1.89 & -14.34&-12.46&0.926 & 0.934 (at 5K)  &  \cite{JCP_1996_104_10008}\\
XII  & 260  & 5000               &  0.77 &    0.87 & 1.40 & 2.27 & -13.23&-10.96 &1.297 & 1.292  & \cite{N_1998_391_00268}  \\
XIII   & 80  & 1                &   0.24 &    0.46 & 1.32 & 1.77 & -13.95&-12.17&1.242 & 1.244  & \cite{S_2006_311_01758} \\
XIV  & 80  & 1                   &  0.24 &    0.46 & 1.34 & 1.80 & -13.99&-12.20&1.307 &   1.332 &\cite{S_2006_311_01758}\\
\hline
\end{tabular}
\end{center}
\end{sidewaystable}

\clearpage
\begin{table}
\begin{center}
\caption{\label{unit_cell}Unit cell parameters for the TIP4PQ/2005 model for a selection of ice phases. Experimental values are from Table 11.2  of Ref \onlinecite{bookPhysIce}. Note that for ice II the hexagonal unit cell, rather than the rhombohedral unit cell,  is given. All distances are in angstroms.}
\begin{tabular}{lccp{0.2cm}llp{0.2cm}ll}
\hline
Phase & $T$ (K)& $p$(bars) && \multicolumn{2}{c}{unit cell} \\ \cline{5-6}
      &      &      &  &  experimental& simulation \\
\hline
I$_h$   &  250 &   0  &&  a=4.518, c=7.356 & a=4.483, c= 7.352\\
II  &  123 &   0  &&  a=12.97, c=6.25 & a= 12.98, c=6.23 \\
III &  250 & 2800 &&  a=6.666, c=6.936 & a=6.645, c=7.011 \\
V   &  100 & 1 &&   a=9.22, b =7.54,            & a=9.06, b=7.64,                \\
    &      &   &&    c=10.35,  $\beta = 109.2^{\mathrm{o}}$ &    c=10.21,  $\beta = 108.6^{\mathrm{o}}$ \\
VI  &  225 & 11000  &&  a=6.181, c=5.698 & a=6.167, c=5.713 \\
\hline
\end{tabular}
\end{center}
\end{table}

\clearpage
\begin{table}
\begin{center}
\caption{\label{table_zero_2}PIMC results for the TIP4PQ/2005 model for the low temperature ice phases at a pressure of 1 bar. All energies are in units of kcal/mol  and the densities are in g$\cdot$cm$^{-3}$.
The errors (in kcal/mol) are ${\mathcal {O}}(0.003)$ in  $K_{\mathrm {translational}}$,  ${\mathcal {O}}(0.02)$ in  $K_{\mathrm {rotational}}$,  
${\mathcal {O}}(0.02)$ in  $U$,  ${\mathcal {O}}(0.04)$ in  $E$ and  ${\mathcal {O}}(0.002)$ g$\cdot$cm$^{-3}$ in  $\rho$ . 
}
\begin{tabular}{llllllll}
\hline
Phase & $T$ (K)  &  $K_{\mathrm {translational}}$ &  $K_{\mathrm {rotational}}$   & $K_{\mathrm {total}}$  &  $U$  &  $E$   & $\rho$ \\
\hline
I$_h$   &  300   &    0.97 &1.49& 2.47& -13.46 &       -10.99    &    0.915 \\
I$_h$   &  200   &    0.71 &1.44& 2.15& -13.98 &       -11.82    &    0.925 \\
I$_h$   &  150   &    0.60 &1.42& 2.02& -14.18 &       -12.16    &    0.928\\
I$_h$   &  125   &    0.54 &1.41& 1.96& -14.25 &       -12.29    &    0.928\\
I$_h$   &  100   &    0.50 &1.42& 1.92& -14.32 &       -12.40    &    0.928\\
I$_h$   &  77    &    0.46 &1.43& 1.89& -14.34 &       -12.45    &    0.927 \\
\hline
II    &  125   &      0.53 &1.32& 1.84& -14.06&        -12.21    &    1.185 \\
II    &  100   &      0.48 &1.30& 1.78& -14.14&        -12.35    &    1.188 \\
II    &  77    &      0.44 &1.32& 1.76& -14.16&        -12.40    &    1.190\\
\hline
III    &  150   &     0.59 &1.36& 1.95& -13.83&        -11.88    &    1.134 \\
III    &  125   &     0.54 &1.36& 1.90& -13.92&        -12.02    &    1.139 \\
III    &  100   &     0.50 &1.37& 1.87& -13.98&        -12.12    &    1.142 \\
III    &  77    &     0.45 &1.37& 1.82& -14.01&        -12.19    &    1.146\\
\hline
V    &  125   &       0.53 &1.34& 1.84& -13.80&        -11.93    &    1.248\\
V    &  100   &       0.49 &1.34& 1.82& -13.88&        -12.06    &    1.251\\
V    &  77    &       0.44 &1.35& 1.79& -13.91&        -12.12    &    1.253\\
\hline
VI    &  125   &      0.53 &1.32& 1.85& -13.67&        -11.82    &    1.330\\
VI    &  100   &      0.48 &1.31& 1.79& -13.74&        -11.95    &    1.334\\
VI    &  77    &      0.45 &1.32& 1.77& -13.77&        -12.00    &    1.336\\
\hline
\end{tabular}
\end{center}
\end{table}

\clearpage
\begin{table}
\begin{center}
\caption{\label{table_pressure_zero_2}Estimates of the coexistence pressures (in bar) for the TIP4PQ/2005 model extrapolated to  0K. Experimental values are taken from the work of Whalley \cite{JCP_1984_81_04087} and the values for the classical TIP4P/2005 model are from \cite{JCP_2007_127_154518}.}
\begin{tabular}{lrrr}
\hline
Phases         & TIP4P/2005 &    TIP4PQ/2005 & Experimental value    \\
\hline
I$_h$-II       & 2090 & 400              &          $ 140 \pm 200    $         \\ 
I$_h$-III      & 3630 & 3008             &           $ 2400 \pm 100   $        \\
II-V           & 11230 & 15630           &             $ 18500 \pm 4000 $       \\
II-VI          & 8530 & 10190            &             $ 10500 \pm 1000 $       \\
III-V          & 3060 & 1800             &             $ 3000 \pm 100   $       \\
V-VI           & 6210 & 5580             &        $ 6200 \pm 200   $       \\
\hline
\end{tabular}
\end{center}
\end{table}

\clearpage

\clearpage

\clearpage

\clearpage

\clearpage

\clearpage

\clearpage

\clearpage

\end{document}